\documentclass[12pt,letterpaper]{article}
\usepackage[usenames, dvipsnames]{xcolor}
\usepackage{jheppub}

\usepackage{graphicx}
\usepackage{subcaption}
\usepackage{bm}
\usepackage{amsmath}
\usepackage{amssymb}
\usepackage{amsthm}

\allowdisplaybreaks[1]

\newtheorem{provisionalRFC}{Provisional Repulsive Force Conjecture}

\newtheoremstyle{named}{}{}{\itshape}{}{\bfseries}{.}{.5em}{#3}
\theoremstyle{named}
\newtheorem*{namedconjecture}{Conjecture}

\newcommand{\cF}{\mathcal{F}}
\newcommand{\df}{\mathrel{:=}}
\newcommand{\noeq}{\mathrel{\phantom{=}}}
\newcommand{\Lgauge}{\Lambda_{\rm gauge}}
\newcommand{\Lphi}{\Lambda_{\phi}}
\newcommand{\Lgrav}{\Lambda_{\rm grav}}
\newcommand{\e}{\mathrm{e}}

\newcommand{\SU}[1]{\mathrm{SU}\!\left(#1\right)}

\definecolor{cobalt}{RGB}{44, 98, 120}
\definecolor{celadon}{rgb}{0.67, 0.88, 0.69}
\definecolor{dm}{cmyk}{.20, 0, .30, 0}
\definecolor{burgundy}{rgb}{0.5, 0.0, 0.13}
\definecolor{plotBlue}{RGB}{94, 130, 181}

\def\be{\begin{equation}}
\def\ee{\end{equation}}

\hypersetup{
  colorlinks,
  citecolor=Violet,
  linkcolor=cobalt,
  urlcolor=Blue}

\begin{document}
\begin{titlepage}

\setcounter{page}{0} \baselineskip=15.5pt \thispagestyle{empty}

\bigskip\

\vspace{2cm}
\begin{center}
{\fontsize{20}{28} \bfseries Repulsive Forces and the \\ \vspace{0.3cm} Weak Gravity Conjecture}
\end{center}
\vspace{1cm}

\begin{center}
\scalebox{0.95}[0.95]{{\fontsize{14}{30}\selectfont Ben Heidenreich,$^{a}$ Matthew Reece,$^{b}$ Tom Rudelius$^{c}$}} 
\end{center}

\begin{center}
\vspace{0.25 cm}
\textsl{$^{a}$Department of Physics, University of Massachusetts, Amherst, MA 01003 USA}\\
\textsl{$^{b}$Department of Physics,  Harvard University, Cambridge, MA 02138 USA}\\
\textsl{$^{c}$Institute for Advanced Study, Princeton, NJ 08540 USA}\\

\vspace{0.25cm}

\end{center}

\vspace{1cm}
\noindent

The Weak Gravity Conjecture is a nontrivial conjecture about quantum gravity that makes sharp, falsifiable predictions which can be checked in a broad range of string theory examples. However, in the presence of massless scalar fields (moduli), there are (at least) two inequivalent forms of the conjecture, one based on charge-to-mass ratios and the other based on long-range forces. We discuss the precise formulations of these two conjectures and the evidence for them, as well as the implications for black holes and for ``strong forms'' of the conjectures. Based on the available evidence, it seems likely that both conjectures are true, suggesting that there is a stronger criterion which encompasses both. We discuss one possibility.

\vspace{1.1cm}

\bigskip
\noindent\today

\end{titlepage}

\setcounter{tocdepth}{2}
\tableofcontents


\section{Introduction}

The Weak Gravity Conjecture (WGC)~\cite{Arkanihamed:2006dz} is most often motivated by a statement about black holes: if all subextremal black holes in a given quantum gravity are kinematically unstable, then conservation of charge and energy imply that there is some charged particle in the spectrum of the theory whose charge-to-mass ratio is at least as large as that of an extremal black hole. The WGC postulates that such a particle exists. This conjecture is intrinsically about gravitational theories, and goes by the slogan ``gravity is the weakest force,'' meaning that gravitational interactions are insufficient to make a stable bound state (the black hole).

However, there is another version of the conjecture, originating in~\cite{Arkanihamed:2006dz} but emphasized more recently by Palti~\cite{Palti:2017elp}: there is a charged particle with the property that two copies of the particle repel each other when they are far apart (a ``self-repulsive'' particle). In other words, the long-range repulsive gauge force between the two identical particles must be at least as strong as the combination of all long-range attractive forces between them. We will call this conjecture (and its generalizations) the ``Repulsive Force Conjecture'' (RFC).

How does the Repulsive Force Conjecture relate to the Weak Gravity Conjecture as formulated in the first paragraph? If we assume that the only long-range forces are gravity and electromagnetism, then the RFC requires a charged particle with charge-to-mass ratio greater than or equal to some critical value (to ensure that the electromagnetic repulsion between two copies is stronger than their gravitational attraction). It is straightforward to check that the long-range force between two extremal Reissner-Nordstr\"om black holes vanishes; therefore, the critical ratio is exactly the charge-to-mass ratio of an extremal Reissner-Nordstr\"om black hole. In other words, the RFC and the WGC are the same conjecture under these assumptions.

Notice, however, that the RFC can be stated without specifically referring to gravity. This is an important distinction, because long range attractive interactions can also be mediated by massless scalar fields. This has two consequences: (1) in quantum gravities with massless scalars, the RFC and the WGC, as defined above, are not identical, and (2) the RFC is also a nontrivial conjecture about \emph{quantum field theories}, since both repulsive (gauge) and attractive (scalar) interactions are possible.\footnote{The quantum field theory RFC is a slight modification of the quantum gravity RFC, see~\S\ref{sec:RFCnongrav}.}

In this paper, we will explore the connection between the WGC and the RFC. In the process, we will fill in many details about the RFC that have not previously appeared in the literature, including formulating a precise definition in theories with multiple gauge bosons. We find that, while neither the WGC nor the RFC implies the other conjecture, violating one while satisfying the other requires physics that seems unlikely to be realized in an actual quantum gravity. For most arguments supporting the WGC, there is a parallel argument supporting the RFC, indicating that \emph{both} conjectures may be true.
This suggests that a
 stronger statement, implying both conjectures, should hold, and we discuss one candidate.

We also explore two interesting generalizations of the RFC. Firstly, strong forms of the WGC such as the Sublattice WGC (sLWGC)~\cite{Heidenreich:2016aqi} and the Tower WGC~\cite{Andriolo:2018lvp} also have self-force analogs, and these conjectures have not been thoroughly explored in previous literature. 
Secondly, as discussed above, the RFC can be generalized to quantum field theories, and we discuss what evidence supports it in these cases, as well as what further calculations could be done to test it.

Note that, since the RFC and WGC collapse to a single conjecture when there are no massless scalar fields, they are essentially identical conjectures in theories without supersymmetry. However, almost all tests of the WGC involve supersymmetry in some way, and thus the distinction can become important. Moreover, comparing these two conjectures leads naturally to slightly stronger conjectures (see~\S\ref{sec:vs}), which remain distinct even without massless scalars.

\bigskip

Before proceeding with our analysis, let us be clear about the history of these ideas, as well as the reasons behind the terminology that we choose in discussing them. All of the topics discussed in this paper fall under the general heading of the ``Weak Gravity Conjecture,'' and both the WGC and the RFC, as we define them, can be traced to ideas in~\cite{Arkanihamed:2006dz}. Subsequently, the WGC version of the conjecture has received more attention, whereas the RFC version was reemphasized by~\cite{Palti:2017elp} and further discussed in \cite{Lust:2017wrl, SimonsTalk, Lee:2018spm}. In particular, \cite{Lee:2018spm} recognized that these conjectures are distinct but argued that they become identical at weak coupling in certain circumstances. We will see similar examples below, but also clarify the logical independence of the conjectures.

These conjectures, therefore, are not new. However, because we wish to carefully distinguish between conjectures based on different (though interrelated) underpinnings, we cannot refer to all of them as the ``Weak Gravity Conjecture.'' Since one conjecture intrinsically involves gravity, whereas the other is about long range forces in general, and does not require gravity, we have chosen the names ``Weak Gravity Conjecture'' (WGC) and ``Repulsive Force Conjecture'' (RFC) to more accurately describe them.

\section{Defining the conjectures}

Our first task is to define carefully what we mean by the ``WGC'' and the ``RFC,'' including the possibility of multiple gauge bosons, massless scalars, etc. We begin with the WGC, which is more familiar and more thoroughly explored in the literature.

\subsection{The Weak Gravity Conjecture (WGC)} \label{subsec:WGC}

To state the conjecture precisely, we will assume a more basic swampland conjecture: the charge $\vec{Q}$ is quantized, i.e., $\vec{Q} \in \Gamma_Q$ for some lattice $\Gamma_Q$ spanning $\vec{Q}$-space \cite{polchinski:2003bq,banks:2010zn,Harlow:2018tng}. A \emph{rational direction} in $\vec{Q}$-space is a ray from the origin which intersects another lattice point. Any nonzero lattice site specifies a unique rational direction, and every rational direction intersects an infinite number of lattice sites with parallel charge vectors. The set of rational directions is dense within the set of all directions (rays from the origin). Central to the conjecture is the charge-to-mass ratio $\vec{Z} \df \vec{Q}/m$ of a massive particle ($m>0$). Because $m$ is not quantized, the physical states of the theory do not form a lattice in $\vec{Z}$-space, but they all lie along rational directions.

\begin{figure}
\centering
\includegraphics[width=6cm]{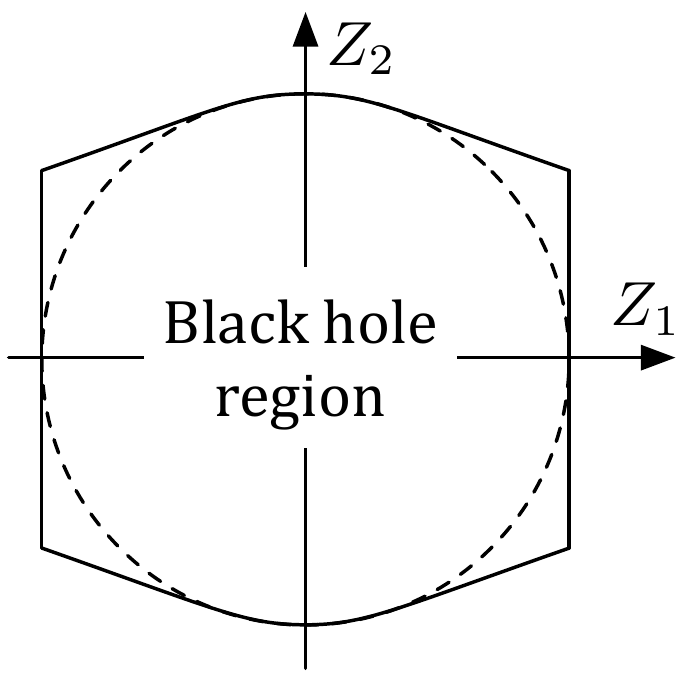}\
\caption{The black hole region in an example with two independent charges. In the presence of massless scalar fields, the black hole region can have an interesting, nontrivial shape, as shown here. If the scalars are given a mass, the region reverts to a ball (contained within the larger region for the original theory), as shown by the dashed line. This particular example is taken from a theory described in \cite{conifolds}, where the straight edges on the left and right correspond to BPS bounds.}
\label{fig:BHregion}
\end{figure}

For a given black hole charge, there is a lower bound on the black hole mass for a semiclassical solution with a horizon to exist (the black hole extremality bound). For parametrically large charge, this lower bound depends only on the two-derivative effective action for the massless fields, and (for vanishing cosmological constant) scales linearly as we scale the magnitude of the charge $\vec{Q} \to \lambda \vec{Q}$. Thus, in the $\lambda \to \infty$ limit, the extremality bound defines a region in $\vec{Z}$-space---the \emph{black hole region} (see Figure~\ref{fig:BHregion})---the interior of which contains parametrically heavy subextremal black holes. Because we took a large charge (and mass) limit, there may not be any black hole states of finite mass (or any states at all) on the boundary of this region, but on the other hand there could be black hole states of finite mass \emph{outside} this region.

The WGC will require that there are states outside or on the boundary of the black hole region. We call such a state \emph{superextremal}; equivalently a superextremal state is a state which \emph{does not lie in the interior} of the black hole region.

To state the full conjecture, it is convenient to formally define a ``multiparticle state'' as consisting of one or more actual particles in the theory,\footnote{By superselection, a particle in the theory (e.g., a single-particle state or a black hole state) must have the same asymptotics as the vacuum.} with ``mass'' $m$ and ``charge'' $\vec{Q}$ equal to the sums of the masses and charges of the constituent particles. 
This corresponds to a limit where the particles in question are taken infinitely far from each other, so that they do not interact. A multiparticle state is superextremal if $\vec{Z} \df \vec{Q}/m$ is outside or on the boundary of the black hole region.

We can now state the Weak Gravity Conjecture in precise terms:
\begin{namedconjecture}[The Weak Gravity Conjecture (WGC)]
For every rational direction in charge space, there is a superextremal multiparticle state.
\end{namedconjecture}
\noindent
When there are a finite number of stable particles in the theory, this is equivalent to the convex hull condition (CHC) of~\cite{Cheung:2014vva}: the convex hull of the stable particles in $\vec{Z}$-space contains the boundary of the black hole region (and thus, its interior as well).

When there are infinitely many (marginally) stable particles we must modify the CHC to a slightly weaker statement: the convex hull of the stable particles in $\vec{Z}$-space contains every rational point along the boundary of the black hole region.\footnote{By ``rational point,'' we mean a point where a rational direction intersects the boundary. The original statement of the CHC is violated by, e.g., maximally supersymmetric theories. In this case the exactly extremal states lie at every rational point along the boundary of the black hole region, but most of the irrational points along the boundary are not contained in the convex hull of these points.} It is then equivalent to the WGC as stated above.

Violating the WGC has interesting consequences for black hole physics. Due to higher derivative operators in the effective action, black holes of finite mass behave differently than parametrically heavy ones, with greater differences for lighter black holes (which have more curvature at their horizon). If the WGC is violated then these corrections must make the lightest black hole of a given finite-but-large charge strictly subextremal. Larger charges (and masses) lead to smaller corrections, so the charge-to-mass ratio of the lightest black hole of a given charge approaches extremality from below as the charge is taken to infinity. Because of the ever-increasing charge-to-mass ratio, the result is an infinite number of stable
 black holes of increasing mass and charge.

This line of reasoning has another interesting consequence: if the convex hull is generated by a finite number of stable particles (the convex hull is ``finitely generated''), then the WGC holds. In particular, we have just shown the contrapositive: if the WGC is violated, then the convex hull \emph{is not} finitely generated. Therefore, a precise formulation of the CHC in the infinitely generated case (as discussed above) is crucial to distinguish spectra that satisfy the WGC from those that violate it; otherwise, the WGC would either be true (if the convex hull is finitely generated) or ambiguous (if it is not).

The WGC may also be extended from particles charged under $1$-form gauge fields to $(p-1)$-branes charged under $p$-form gauge fields.
For $p > 1$, the above statements carry over, with superextremality defined relative to an extremal black brane rather than an extremal black hole, and the charge-to-mass vector $\vec{Z}$ replaced by a charge-to-tension vector $
\vec{Z} \df \vec{Q}/T$.

Although the WGC has been extended to $AdS$ spacetimes in as few as three dimensions \cite{Montero:2016tif}, the conjecture in its most basic form applies only to theories in asymptotically-flat spacetimes in $D \geq 4$ dimensions. In flat space in three dimensions, gravity does not have any propagating degrees of freedom, and massive particles backreact on the spacetime geometry by introducing a deficit angle, which prevents asymptotic flatness. We therefore follow the typical convention and restrict our discussion of the WGC in this paper to the case of $D \geq 4$.

\subsection{The Repulsive Force Conjecture (RFC)} \label{subsec:RFC}

We now develop the RFC using the same principles as the WGC
but with the notion of ``superextremal'' replaced with that of ``self-repulsive.'' After specifying precisely what a ``self-repulsive'' particle is, we  
develop the conjecture for the case of multiple photons. As in the case of the WGC, to avoid the issue of deficit angles, we restrict our discussion in this paper to theories in asymptotically-flat spacetimes in $D \geq 4$ dimensions, though it might be interesting to extend the RFC to theories in fewer dimensions as well.

The force between two massive particles separated by a distance $r$ in $D$ dimensions with vanishing cosmological constant takes the general form:
\begin{equation}
F_{12} = \frac{k^{a b} Q_{1a} Q_{2b}}{r^{D-2}} - \frac{G_N m_1 m_2}{r^{D-2}} - \frac{g^{i j} \mu_{1i} \mu_{2j}}{r^{D-2}}+\ldots \,, \label{eqn:longrangeforce}
\end{equation}
in the large $r$ limit, where $Q_a$ are the gauge charges, $\mu_i$ are the scalar ``charges,'' we suppress vector notation for simplicity and $F_{12}>0$ ($F_{12}<0$) corresponds to a repulsive (attractive) force.  Here we assume that the deep infrared is described by the Einstein-Hilbert action coupled to gauge bosons and neutral, massless scalars; this assumption allows us to ignore logarithmic factors that could arise in the presence of massless charged particles.

The leading-order ``long range'' force falls off like $1/r^{D-2}$, with contributions from massless spin one, spin two, and spin zero bosons---the three terms in~(\ref{eqn:longrangeforce}). We refer to all contributions falling off more quickly than this as ``short-range.'' Writing $F_{I J} = \frac{\mathcal{F}_{I J}}{V_{D-2} r^{D-2}} + \ldots$ for any two partices $I$ and $J$,\footnote{We include a factor of the $(D-2)$-sphere volume $V_{D-2}$ in the definition of $\mathcal{F}_{I J}$ for future convenience.} we say that $I$ and $J$ are \emph{mutually repulsive} if the mutual-force coefficient $\mathcal{F}_{I J}$ is non-negative, and that $I$ is \emph{self-repulsive} if the self-force coefficient $\mathcal{F}_{I I}$ is non-negative.

In particular, BPS states are ``self-repulsive'' due to the stronger condition $F_{I I} = 0$ (the force between identical BPS states is zero). 
 One might worry that we are mislabeling particles with $\mathcal{F}_{I I} = 0$ but $F_{I I} < 0$ (i.e., those for which the long-range force vanishes while the short-range force is attractive) as ``self-repulsive.'' 
The reason for this particular choice is explained below.
Note, however, that it is highly unlikely for $\mathcal{F}_{I I}$ to vanish exactly unless $I$ is a BPS state, so this exceptional case probably never occurs in real examples (cf.~\cite{Ooguri:2016pdq}). 

The significance of self-repulsiveness is especially pronounced in four dimensions. This is best illustrated by considering its opposite case: a \emph{self-attractive} particle is one with $\mathcal{F}_{I I}<0$. A (massive) self-attractive particle in four dimensions can form a bound state with itself with strictly negative binding energy. By comparison, if $F_{I I} < 0$ but $\mathcal{F}_{II} = 0$, the existence of a bound state with negative binding energy depends on the details of the short-range forces. This is why we count this case as ``self-repulsive'': a bound state is not guaranteed.

Consider a theory in four dimensions with a single massless photon and no self-repulsive particles, and assume for simplicity that all charged particles are massive. By assumption, any charged particle in the theory is massive and self-attractive. The bound state of two copies of the particle is either stable---in which case it is a new particle species with larger charge-to-mass ratio than the original---or it decays to some combination of stable particles, one of which must have higher charge-to-mass ratio than the original because of the strictly negative binding energy. Iterating this procedure, we conclude that the theory contains an infinite number of stable charged particle species with increasingly large charge-to-mass ratios~\cite{Arkanihamed:2006dz} (assuming the theory contains any charged particles at all, which it must to obey more general ``no global symmetries'' arguments \cite{polchinski:2003bq,banks:2010zn,Harlow:2018tng}). 

This tower of states is very similar to the tower of states in a theory that violates the WGC, but now instead of near-extremal black holes the stable states originate from weakly bound states under the long range forces, or their decay products. Thus, by analogy with the claim that a quantum gravity with a massless photon contains a superextremal particle~\cite{Arkanihamed:2006dz}, we conjecture:
\begin{provisionalRFC} \label{conj:SRparticle}
For any massless photon in a quantum gravity, there is a self-repulsive particle charged under the photon.~\cite{Palti:2017elp}
\end{provisionalRFC}
\noindent
This is the conjecture formulated by Palti, and we consider it to be foundational in defining what is meant by the ``repulsive force conjecture.'' However, the conjecture and the motivation that led us to it come with several important subtleties that must be addressed.

First, although the conjecture makes sense in any number of dimensions, in motivating it we were careful to restrict ourselves to $D=4$. In $D > 4$ dimensions the consequences of self-attractiveness are not so simple. As discussed in appendix \ref{sec:bound}, although \emph{classically} $\mathcal{F}_{I I}<0$ is sufficient to ensure a bound state in any dimension,
this is not true quantum-mechanically in $D>4$ dimensions. Thus, an Abelian gauge theory with only self-attractive charged particles does not necessarily produce an infinite tower of stable charged states with increasing charge-to-mass ratios. Nonetheless, it is possible to motivate the conjecture in higher dimensions by compactifying to $D=4$. We will return to this point in \S\ref{ssec:RFCdimred}.

Second, the notion of ``repulsiveness'' is not well-defined for massless charged particles. To deal with this issue, we formally extend the right-hand side of (\ref{eqn:longrangeforce}) to the $m_i = 0$ case and declare two particles to be mutually repulsive (attractive) if $\mathcal{F}_{12} \geq 0$ ($< 0$). As a consequence, a bound state is not guaranteed between two mutually attractive particles if at least one of them is massless, even in four dimensions. Note that, while in the absence of massless scalars massless charged particles are necessarily self-repulsive, this is no longer guaranteed in the presence of long-range scalar forces. 

Third, it is not immediately clear from the definition above what type of ``particle'' is allowed to satisfy the conjecture: must the particle be stable, or can it be a long-lived, unstable resonance? In the case of the WGC, this question was irrelevant because conservation of charge and energy imply that a charged resonance can only decay to a multiparticle state with a charge-to-mass ratio at least as large as that of the original resonance.
 Thus, a superextremal resonance will always decay to a superextremal multiparticle state.
However, in part because the scalar ``charge'' $\mu$ is in general \emph{not conserved}, in the presence of scalar forces there is no guarantee that the decay of a self-repulsive resonance will produce any self-repulsive particles.\footnote{As explained in the text below conjecture~\ref{conj:WSRmultiparticle}, even conservation of scalar charge would not guarantee a self-repulsive particle in the final state, because a multiparticle state can be ``on-average'' self-repulsive without containing any self-repulsive particles.} In principle, this means that self-repulsive resonances can exist without there being any self-repulsive stable particles in the spectrum.

For the purposes of this paper, we will allow long-lived, unstable particles to satisfy the requirements of the repulsive force conjecture. This choice comes with drawbacks and advantages. The downside is a lack of precision: the conjecture is sharply defined only when the theory is parametrically weakly coupled, since otherwise the definition of a ``resonance'' becomes unclear.
The upside is significant practical gain: the question of whether or not a theory contains a self-repulsive (possibly unstable) particle can typically be addressed simply by considering the tree-level spectrum. Ensuring that such a self-repulsive particle is stable, on the other hand, requires detailed knowledge about the spectrum of bound states in the theory, which makes it very difficult to check that the repulsive force conjecture is satisfied by stable particles alone. Furthermore, allowing for unstable resonances significantly simplifies the discussion of strong forms of the RFC, as we will see in \S\ref{ssec:RFCSTRONG}.

Finally, the self-repulsive particle guaranteed by the provisional conjecture we have defined above may owe its self-repulsion primarily to a different gauge field in the theory, so it reduces to a weaker statement than the convex hull condition when there are multiple photons but no massless scalars. This was already pointed out by Palti~\cite{Palti:2017elp}. We  solve this problem below by formulating a stronger conjecture.

As in the previous section, it is convenient to consider formal multiparticle states (which may include long-lived unstable particles, as discussed). Unlike before, there are multiple notions of self-repulsiveness that are interesting to consider. We say that a multiparticle state is \emph{weakly} (or ``on-average'') self-repulsive if the \emph{total} mass, charge and scalar charge of the state (defined as the sum of the masses, charges, and scalar charges of the constituents) leads to self-repulsion. Likewise, a multiparticle state is \emph{strongly} (or ``in-detail'') self-repulsive if any two (not necessarily distinct) particles in the state are mutually repulsive.

In other words, letting $n^I$ denote the number of particles of species $I$ (counting antiparticles as a different species) in the multiparticle state, the state is \emph{weakly (on-average) self-repulsive} if $\sum_{I, J} n^I n^J \mathcal{F}_{I J} \ge 0$ and \emph{strongly (in-detail) self-repulsive} if $\mathcal{F}_{I J} \ge 0$ for all $I,J$ in the multiparticle state. Clearly a strongly self-repulsive state is weakly self-repulsive. 

A seemingly straightforward analog of the convex hull condition is
\begin{provisionalRFC} \label{conj:WSRmultiparticle}
For every rational direction in charge space, there is a weakly self-repulsive multiparticle state.
\end{provisionalRFC}
\noindent
However, this conjecture is too weak to be very interesting. In particular, it does not even imply conjecture~\ref{conj:SRparticle}! A multiparticle state consisting entirely of self-attractive particles can nonetheless be weakly self-repulsive. Consider, for example, two particles with equal mass $m$ and charge $Q$, but opposite scalar charge $\mu_2 = - \mu_1$. The multiparticle state is weakly self-repulsive so long as $k^2 Q^2 \ge G_N m^2$, but for large enough $|\mu_1| = |\mu_2|$, both constituents can be made self-attractive.

One solution to this problem---in some sense combining conjectures~\ref{conj:SRparticle} and~\ref{conj:WSRmultiparticle}---is to formulate the stronger conjecture
\begin{provisionalRFC} \label{conj:WSRmultiparticleSRparticle}
For every rational direction in charge space, there is a weakly self-repulsive multiparticle state consisting entirely of self-repulsive particles.
\end{provisionalRFC}
\noindent
Now it is obvious that self-repulsive particles must exist to satisfy the conjecture. In particular, this implies both conjecture~\ref{conj:SRparticle} and conjecture~\ref{conj:WSRmultiparticle}.

\begin{figure}
\centering
\includegraphics[width=6cm]{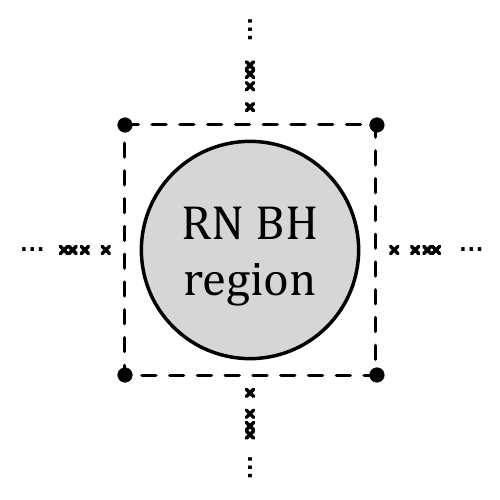}\
\caption{A spectrum for a 4d theory that violates the RFC but satisfies conjecture~\ref{conj:WSRmultiparticleSRparticle}, plotted in $\vec{Z}$ space with the Reissner-Nordstr\"om black hole region shown. The four black dots represent self-repulsive particles with no scalar couplings; each distinct pair is mutually attractive, so these particles satisfy conjecture~\ref{conj:WSRmultiparticleSRparticle} but not the RFC. The bound states of the adjacent pairs are the four innermost crosses in the diagram. The unknown scalar couplings of these bound states can be assumed to be large enough that they are self-attractive, with opposite signs for adjacent pairs, so that they are mutually repulsive. Each one binds to itself and (assuming the bound states remain self-attractive) produces an infinite tower of bound states, represented by a line of crosses in the diagram. With appropriate assumptions about the details, the spectrum as pictured is complete.}
\label{fig:conjecture3vsRFC}
\end{figure}

Conjecture~\ref{conj:WSRmultiparticleSRparticle} has some of the properties that we want from the repulsive force conjecture. However, as we will see, this conjecture still allows spectra with many of the same characteristics as those violating conjecture~\ref{conj:SRparticle}. We instead focus on a simpler and even stronger conjecture as our working definition of the RFC:
\begin{namedconjecture}[The Repulsive Force Conjecture (RFC)]
For every rational direction in charge space, there is a strongly self-repulsive multiparticle state.
\end{namedconjecture}
\noindent
This implies conjectures~\ref{conj:SRparticle}, \ref{conj:WSRmultiparticle}, and \ref{conj:WSRmultiparticleSRparticle}. On the other hand, it is not difficult to devise spectra in $D >5$ dimensions that violate the RFC but satisfy conjecture~\ref{conj:WSRmultiparticleSRparticle} (and therefore conjectures~\ref{conj:SRparticle} and~\ref{conj:WSRmultiparticle} as well). In 4d, mutually attractive particles necessarily form bound states, so the spectrum must be ``complete'': whenever two particles are mutually attractive, either their bound state is itself a stable particle in the spectrum, or there is a multiparticle state in the spectrum to which it can decay. One example of a complete spectrum satisfying conjecture~\ref{conj:WSRmultiparticleSRparticle} but violating the RFC is shown in Figure~\ref{fig:conjecture3vsRFC}. Infinite towers of weakly bound states appear, a common characteristic of RFC-violating spectra in four spacetime dimensions.

This example demonstrates that consistency of the low-energy effective field theory alone does not ensure that a theory satisfying conjecture~\ref{conj:WSRmultiparticleSRparticle} must also satisfy the RFC, even in four dimensions. However, the spectrum is contrived and we do not expect it to be realized in a UV-complete theory of quantum gravity.
Indeed, it is possible that all violations of any one of the above conjectures are confined to the Swampland.

Let us see what happens when the RFC, as just formulated, is violated in four dimensions. A ``weakly self-attractive'' multiparticle state is one that is not strongly self-repulsive, i.e., $\mathcal{F}_{I J} < 0$ for some $I, J$ in the state. If this holds for some $I \ne J$,  
then in four dimensions we obtain a new multiparticle state with less mass and the same charge by replacing $I$ and $J$ with their bound state, or its decay products. Otherwise, $\mathcal{F}_{I I} < 0$ for some particle $I$ in the state, and by combining \emph{two copies} of the original multiparticle state and then replacing $I$ and its duplicate $I'$ with their bound state or its decay products, we obtain another multiparticle state with twice the charge and less than twice the mass (hence a larger charge-to-mass ratio, as before). 

If the RFC is violated in four dimensions, then at least one rational direction in charge space has no strongly self-repulsive multiparticle states along it. Pick any multiparticle state along this direction,\footnote{Such a multiparticle state is guaranteed to exist so long as each gauge boson couples to at least one charged particle.} which is weakly self-attractive by assumption. As explained above, we can obtain from this multiparticle state another one with a parallel charge vector and strictly larger charge-to-mass ratio. Iterating this procedure, we find an infinite tower of multiparticle states with ever increasing charge-to-mass ratios. Violating any of the weaker conjectures discussed above has the same consequence.

Note that, if the convex hull of stable particles in $\vec{Z}=\vec{Q}/m$ space is finitely generated, then for every rational direction in charge space there is a multiparticle state of maximum $|\vec{Z}|$.\footnote{We say that such a theory satisfies the ``Maximal Z Conjecture,'' see~\S\ref{sec:MZC}.} In particular, this means that a finitely generated convex hull implies the RFC in four dimensions, just as it implies the WGC in any dimension.

The RFC and WGC are closely related. Without massless scalar fields, the third term in (\ref{eqn:longrangeforce}) is absent, and self-repulsiveness is determined by  charge-to-mass ratio. One can check that extremal black holes in these (two-derivative) Einstein-Maxwell theories (i.e., $D$-dimensional extremal Reissner-Nordstr\"om solutions~\cite{myers:1986un}) have zero self-force and so self-repulsive and superextremal single-particle states are the same. 

However, when there are multiple photons this does not quite make the RFC and the WGC equivalent. In particular, without massless scalars a superextremal multiparticle state is the same as a weakly self-repulsive multiparticle state, making conjecture~\ref{conj:WSRmultiparticle} manifestly equivalent to the WGC. Since the RFC implies conjecture~\ref{conj:WSRmultiparticle}, we conclude that the RFC implies the WGC in this context. On the other hand, the converse is far from obvious: because a superextremal multiparticle state is not necessarily strongly self-repulsive, the RFC may be stronger than the WGC in the presence of multiple photons and no massless scalars.

Indeed, in $D>4$ dimensions we can easily write down spectra which satisfy the WGC and violate the RFC. However, these spectra are typically ``incomplete'': they contain pairs of mutually attractive particles with no corresponding bound state or bound state decay products in the spectrum, which renders them inconsistent in 4d. Thus, to show that the RFC follows from the WGC in 4d, we would have to leverage this completeness requirement. At present, we do not know an argument that does so, but likewise it is very difficult to write down a complete spectrum that satisfies the WGC and not the RFC. 

\begin{figure}[t!]
    \centering
    \includegraphics[width = 100mm]{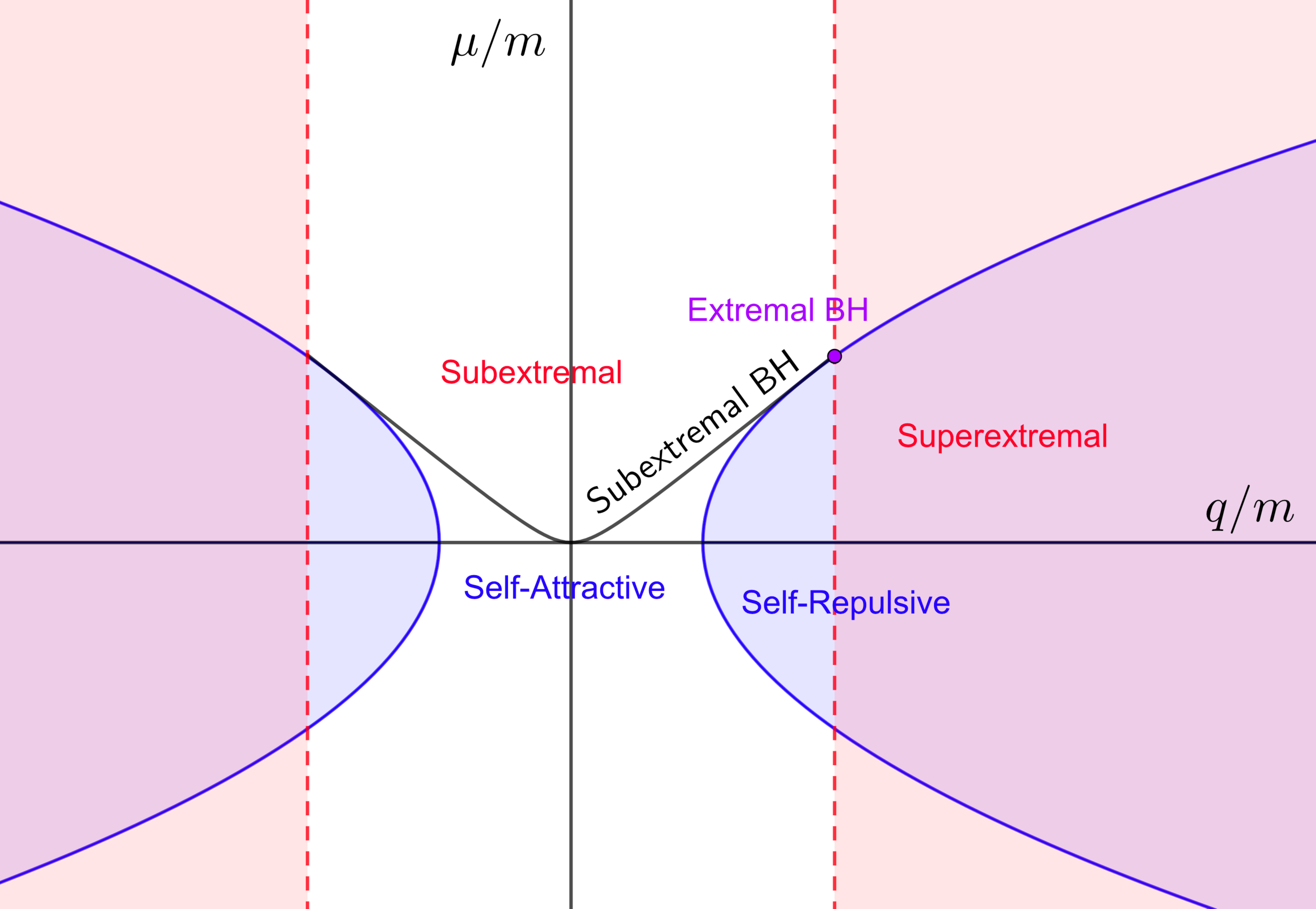}
    \caption{Superextremality regions (shaded red) and self-repulsiveness regions (shaded blue) in gauge charge-to-mass and scalar coupling-to-mass space. States with larger $\mu/m$ than an extremal black hole can be superextremal but not self-repulsive, while states with smaller $\mu/m$ than an extremal black hole can be self-repulsive but not superextremal. }
    \label{fig:phase}
\end{figure}

In theories with massless scalar fields, 
neither the WGC nor the RFC implies the other conjecture.
It is still the case that extremal black holes have vanishing self-force \cite{BHpaper}. However, charged particles may couple differently to the moduli than extremal black holes do. A superextremal particle which couples more strongly to moduli than the corresponding black hole can be self-attractive, and likewise a subextremal particle which couples more weakly to the moduli than the corresponding black hole can be self-repulsive.
These various possibilities are illustrated in figure \ref{fig:phase}. Further comparisons between the WGC and RFC are discussed in~\S\ref{sec:vs}.

\section{Review of the evidence for the WGC}

A number of lines of evidence have been provided in favor of the WGC. Before proceeding with our analysis of the RFC, we review some of them, focusing on 1) dimensional reduction, 2) modular invariance, 3) examples in string theory, 4) gauge-gravity unification, 5) infrared consistency, and 6) various black hole arguments.

\subsection{Dimensional reduction}\label{eq:dimredWGC}

If the WGC holds in any quantum gravity then it must remain true after compactification on a circle. It turns out that if we ignore the Kaluza-Klein photon then in general the WGC in the higher dimensional theory implies the WGC in the lower dimensional theory. (We return to the question of the KK photon in~\S\ref{subsec:RFCKKmodes} and~\S\ref{sec:STRONG}, motivating strong forms of both the WGC and the RFC.)

This computation was first carried out in \cite{Heidenreich:2015nta}, but we review it here.  We begin with an Einstein-Maxwell-dilaton action for a $P$-form in $D= d+1$ dimensions,
\be
S = \frac{1}{2\kappa_D^2} \int d^D x \sqrt{-g} \left({\cal R}_D - \frac{1}{2} (\nabla \phi)^2\right) - \frac{1}{2e_{P;D}^2} \int d^D x \sqrt{-g} \e^{-\alpha_{P;D} \phi} F_{P+1}^2 \,. \label{eq:generalaction}
\ee
Here $F_{P+1} = d A_P$ is the field strength for a $P$-form gauge field $A_{\mu_1 \ldots \mu_P}$, with
\be
F_q^2 \df \frac{1}{q!} F_{\mu_1 \ldots \mu_q} F^{\mu_1 \ldots \mu_q} \,.
\ee
We use the convention
\be
\frac{1}{\kappa_D^2} = M_D^{D-2},
\ee
with $M_D$ the reduced Planck mass in $D$ dimensions. With this convention, the condition for a $(P-1)$-brane of quantized charge $q$ and tension $T_P$ to be superextremal is given by:
\begin{equation}
    e_{P;D}^2 q^2 M_{D}^{D-2} \geq \left[ \frac{\alpha_{P;D}^2}{2} + \frac{P(D-P-2)}{D-2} \right] T_P^2.
    \label{eq:extremalitybound}
\end{equation}
We will sometimes refer to this inequality as the ``WGC bound.''

We consider a dimensional reduction ansatz of the form,
\be
ds^2 = \e^{\frac{\lambda(x)}{d-2}} d{\hat s}^2(x) + \e^{-\lambda(x)} dy^2,
   \label{eq:dimredansatz}
\ee
where $y \cong y + 2 \pi R$. This ansatz is chosen so that the dimensionally reduced action is in Einstein frame, i.e., it eliminates the kinetic mixing between $\lambda$ and the $d$-dimensional metric. Note that we are not yet including a Kaluza-Klein photon, but we do include a massless radion, which controls the radius of the circle. Under such a dimensional reduction, the scalar metric and gravitational constant change according to:
\begin{align}
G_{ij}^{(d)} &= (2 \pi R) G_{ij}^{(D)} ,  & 
M_d^{d-2} &= (2 \pi R) M_D^{D-2} . \label{eq:Ddconstants}
\end{align}
In the remainder of this section we will assume the asymptotic behavior $\lambda \to 0$ as $x \to \infty$; in later sections, when we will be interested in the derivatives of various quantities with respect to the vacuum expectation value of $\lambda$, we will relax this assumption. 

The $P$-form in $D$ dimensions gives both a $P$-form and a $p$-form in $d$ dimensions, with $p=P-1$. The former comes from taking all of the legs of the $P$-form to lie along noncompact directions, while the latter comes from taking one of the legs of the $P$-form to lie along the compact circle. Likewise, a $(P-1)$-brane charged under the $P$-form descends to both a $(P-1)$-brane and a $(p-1)$-brane, charged under the respective forms. We consider these two cases in turn.

\subsubsection{$P$ preserved}

We begin with the $P$-form in $d$ dimensions. The associated gauge coupling is given by
\begin{equation}
    e_{P;d}^2 = e_{P;D}^2/(2 \pi R).
\end{equation}
The tension of a $(P-1)$-brane transverse to the compact circle is unchanged, $T_{P;d} =T_{P;D} \df T_P$.

After reduction, the radion $\lambda$ and dilaton $\phi$ each couple exponentially to the Maxwell term in the action. We can therefore redefine the dilaton to absorb the coupling of the radion to the scalar field, which effectively shifts the coupling $\alpha$ appearing in the WGC bound, so that
\begin{equation}
    \alpha_{P;d}^2 = \alpha_{P;D}^2 + \frac{2 P^2}{(d-1)(d-2)}.
\end{equation}
Plugging this into the WGC bound (\ref{eq:extremalitybound}) with $D \rightarrow d$, we have
\begin{align}
    e_{P;d}^2 q^2 M_{d}^{d-2} \geq \left[ \frac{\alpha_{P;d}^2}{2} + \frac{P(d-P-2)}{d-2} \right] T_{P}^2
    = \left[ \frac{\alpha_{P;D}^2}{2} + \frac{P(D-P-2)}{D-2} \right] T_{P}^2.
\end{align}
We see that the factor appearing on the right-hand side is precisely the factor that appeared in the $D$-dimensional WGC bound. This shows that superextremality is exactly preserved: the $(P-1)$-brane in $d$ dimensions is superextremal if and only if the $(P-1)$-brane in $D$ dimensions was superextremal. If we had instead stabilized the radion, so that it no longer contributes to the extremality bound, we would have found a strictly weaker WGC bound in $d$ dimensions. In either case, satisfying the bound in the parent theory is sufficient to satisfy it in the daughter theory.

\subsubsection{$P$ reduced}

We next consider the $p$-form in $d$ dimensions. The associated gauge coupling is given by
\begin{equation}
    e_{p;d}^2 = (2 \pi R) e_{P;D}^2.
\end{equation}
Wrapping a $(P-1)$-brane on the compact circle gives a $(p-1)$ brane with tension
\begin{equation}
    T_{p;d} = (2 \pi R) T_{P;D}.
\end{equation}
The coupling of the scalar fields to the Maxwell term is slightly different than in the previous case, so we get a new relation for the coupling constant $\alpha$,
\begin{equation}
    \alpha_{p;d}^2 = \alpha_{P;D}^2 + \frac{2 (d-p-2)^2}{(d-1)(d-2)}.
\end{equation}
The WGC bound in $d$ dimensions thus becomes,
\begin{align}
    e_{p;d}^2 q^2 M_{d}^{d-2} \geq \left[ \frac{\alpha_{p;d}^2}{2} + \frac{p(d-p-2)}{d-2} \right] T_{p;d}^2 
    = \left[ \frac{\alpha_{P;D}^2}{2} + \frac{P(D-P-2)}{D-2} \right] T_{p;d}^2.
\end{align}
Once again, this is precisely the factor that appeared in the $D$-dimensional WGC bound. Just as in the previous case, the WGC constraint is exactly preserved if the radion is massless, whereas it is weakened if the radion is stabilized. 

\subsection{Modular invariance}\label{ssec:modular}

In \cite{Heidenreich:2016aqi, Montero:2016tif} (see also \cite{Lee:2018spm, Aalsma:2019ryi}), it was noted that modular invariance of 2d CFTs implies the existence of a sublattice of the same dimension as the full charge lattice in which every site contains a superextremal (i.e., WGC-satisfying) charged particle. These charged particles exist in the NS-NS sector of string theory or in AdS$_3$, depending on whether one views the 2d CFT as a worldsheet theory or a holographic dual. (See \cite{Lin:2019kpn} for further discussion and caveats regarding this latter interpretation.)

In particular, this argument suggests a strong form of the WGC, see~\S\ref{sec:STRONG}.

\subsection{More examples in string theory}

Aside from the cases discussed above, the WGC has been verified in many other examples in string theory. In \cite{Heidenreich:2016aqi}, the WGC was checked in a large number of type II string orbifolds and a handful of holographic CFTs. The existence of BPS states satisfying the WGC in Calabi-Yau compactifications has been discussed in \cite{Grimm:2018ohb, conifolds, Font:2019cxq}, and tests of the WGC in 6d and 4d Calabi-Yau compactifications of F-theory have been carried out in \cite{Lee:2018urn, Lee:2018spm, Lee:2019tst, Lee:2019xtm}. As in the previous subsection, these examples feature infinite towers of superextremal particles, as suggested also by the Swampland Distance Conjecture \cite{Ooguri:2006in}.

\subsection{Gauge-gravity unification}\label{ssec:gaugegrav}

The WGC may also be related to the idea of emergence from an ultraviolet cutoff \cite{Harlow:2015lma, Heidenreich:2017sim}. In particular, let us assume that a weakly-coupled $U(1)$ gauge theory emerges in the IR upon integrating out a tower of charged states below some energy scale $\Lgauge$, at which point the gauge theory loop expansion breaks down. To be more precise, $\Lgauge$ is the scale at which 1PI loop effects from particles of mass below $\Lgauge$ rival tree-level contributions to the propagator, and the parameter
\begin{equation}
\lambda_{{\rm gauge}}( \Lambda) \df e^2 \Lambda^{D-4}  \sum_{i | m_i < \Lambda} q_i^2
\label{eq:Landaupole}
\end{equation}
is equal to 1, i.e., $\lambda_{{\rm gauge}}( \Lgauge) \df 1$.

Likewise, we define $\Lgrav$ to be the scale at which gravity becomes strongly coupled, i.e., the loop effects from particles of mass below $\Lgrav$ rival tree-level contributions to the propagator, and 
\begin{equation}
\lambda_{{\rm grav}}( \Lambda) \df \sum_{i | m_i < \Lambda} (\Lambda/M_{D})^{D-2},
\end{equation}
is equal to 1, i.e., $\lambda_{{\rm grav}}( \Lgrav) \df 1$. If we then assume that gauge theory ``unifies" with gravity in the sense that $\Lgauge \approx \Lgrav$, we find (under certain regularity assumptions on the spectrum) that the average particle in the tower is extremal,
\begin{equation}
e^2 \frac{\langle q^2 \rangle_{\Lgauge}}{\Lgauge^2} \sim \frac{1}{M_D^{D-2}}, \label{eqn:gaugegrav}
\end{equation}
up to order-one factors, where $\langle q^2 \rangle_\Lambda$ is the average charge $q^2$ of particles with mass below $\Lambda$.

\subsection{Infrared consistency}

Integrating out a massive charged particle introduces higher-dimension operators to the effective action. In four dimensions, these take the schematic form $F^4$, $F^2 R$, and their induced coefficients are proportional to powers of the particle's charge-to-mass ratio, $Z$. Unitarity, analyticity, and causality constrain the coefficients of these operators, so if one assumes that the induced terms are the dominant ones in the effective action, this in turn translates to the constraint $Z \geq 1$, i.e., the WGC must be satisfied \cite{cheung:2014ega}. This conclusion carries over to theories with multiple Abelian gauge fields, and in the (seemingly unlikely) case that the only charged fields are scalars, a similar argument implies an infinite tower of superextremal scalar fields \cite{Andriolo:2018lvp}. The assumption that the terms induced from integrating out light fields dominate is likely to be only approximately true, as cutoff-suppressed operators will compete with these terms, in which case the constraint can be relaxed. However, for parametrically-large black holes in theories without massless scalars, the logarithmic running of the $F^4$ coefficient will dominate, ensuring the existence of at least one state (perhaps a large black hole) with $Z \geq 1$ \cite{asymptotic}. For other claims linking the WGC to unitarity, analyticity, and causality, see \cite{Hamada:2018dde, Chen:2019qvr, Bellazzini:2019xts, Mirbabayi:2019iae}.

\subsection{Black hole arguments and cosmic censorship}

A number of papers have recently argued that consistency of black hole physics requires a superextremal state \cite{Hod:2017uqc, Fisher:2017dbc, Cheung:2018cwt, Montero:2018fns} (see however \cite{Cottrell:2016bty, Shiu:2017toy}). We will not attempt to evaluate these arguments or summarize them here. An intriguing set of calculations in classical GR shows that theories in AdS$_4$ that violate the WGC can also violate cosmic censorship \cite{Crisford:2017zpi, Crisford:2017gsb}. Interestingly, in the case of theories with a massless dilaton, the existing calculations support the conjecture we refer to as the WGC (rather than the RFC) as the precise condition needed to avoid violations of cosmic censorship \cite{Horowitz:2019eum}. 

\section{The RFC: basic consistency checks} \label{subsec:RFCbasic}

We have reviewed several lines of evidence in the existing literature in favor of the WGC. Comparatively little has been done to establish the RFC. However, we now show that many of the above lines of evidence have self-force analogs, thereby providing evidence in favor of the RFC. 

\subsection{Computing forces} \label{subsec:forces}

Consider two parallel $(P-1)$ branes in a flat $D$-dimensional background, separated by a parametrically large distance $r$. The branes exert forces on each other mediated by gravitons, gauge fields, and massless scalars. Before proceeding with our analysis of the RFC, we write down a general expression for the leading, long-range force between the branes as a function of their charge, tension, and scalar charge. A derivation is given in~\cite{BHpaper}.

For simplicity, we focus on branes with a Poincar\'e-invariant worldvolume, i.e., with equal energy density $\mathcal{M}$ and tension $\mathcal{T}$, saturating the null-energy condition constraint $\mathcal{T} \le \mathcal{M}$. These are familiar from string theory examples, and can be described by the well-known Dirac membrane action. By comparison, it turns out that \emph{sub-extremal} black branes have $\mathcal{T} < \mathcal{M}$, and are therefore not of this type. 
 We will not discuss such branes in detail in the present paper, but merely quote results from~\cite{BHpaper} where appropriate.\footnote{Note that branes of this type necessarily have additional degrees of freedom relative to Dirac branes, and in particular they typically carry a nonvanishing entropy density. Thus, while Dirac branes are closely analogous to fundamental particles, branes with $\mathcal{T} < \mathcal{M}$ are not.} Unless otherwise stated, all branes discussed below are assumed to be of this simple, boost-invariant ($\mathcal{T} = \mathcal{M}$) type.

The low-energy effective action for the massless fields is of the form,
\begin{equation}
S_D = \int d^Dx \sqrt{-g} \left[\frac{1}{2 \kappa_D^2} \mathcal{R} - \frac{1}{2} G_{ij} \nabla \phi^i \nabla \phi^j - \frac{1}{2 (P+1)!} \tau_{ab} (F_{\mu_1 \mu_2 ... \mu_{P+1}}^a  F^{b,\mu_1 \mu_2 ...\mu_{P+1}})  \right].
   \label{eq:Ddimaction}
\end{equation}
Here we assume for definiteness that the scalar potential vanishes, $V(\phi) = 0$, i.e., the massless scalars are moduli. We work in Einstein frame, so that $\kappa_D^2$ is independent of the moduli $\phi^i$, whereas $G_{i j}(\phi)$ and $\tau_{a b}(\phi)$ can both depend on the moduli. Two $(P-1)$-branes with respective charges $q_{1a}, q_{2a}$ under the gauge group $a$ and tensions $T_1, T_2$ exert a pressure on each other of the form
\begin{align} P_{12} = \frac{\mathcal{P}_{12}}{r^{D - P - 1} V_{D - P - 1}} + \ldots, 
\end{align}
up to subleading terms in the large $r$ limit, where (see, e.g.,~\cite{Lee:2018spm, BHpaper})
\begin{align}
   \mathcal{P}_{12} = \tau^{a b} q_{1 a} q_{2 b} - G^{ij} (\partial_i T_1) (\partial_j T_2) - \frac{P(D-P-2)}{D-2} \frac{T_1 T_2}{M_D^{D-2}},
     \label{eq:repulsion}
   \end{align}
and $\partial_i T \df \frac{\partial T}{\partial \phi^i}$ is the partial derivative of the brane tension with respect to the modulus, holding the Planck scale fixed.\footnote{The same result applies to massless scalars whose potential is nonzero at higher order. Even though $\phi^i \ne 0$ is no longer a vacuum of the theory, $T(\phi)$ can still be interpreted (in a slightly less sharp fashion) as the tension of the brane as a function of the scalar field.} The two branes are mutually repulsive if $\mathcal{P}_{12} \ge 0$ and mutually attractive otherwise. We  mostly  consider the self-force case $q_1 = q_2, T_1 = T_2$, but the general case is relevant for checking whether multiparticle states are strongly self-repulsive when considering the RFC with multiple gauge fields.

\subsection{Dimensional reduction}\label{ssec:RFCdimred}

As in the case of the WGC, we first check how the RFC behaves under dimensional reduction. To begin with, we reduce a $P$-form Einstein-Maxwell-dilaton theory on a circle, both reducing and preserving $P$. We account for the radion in our analysis, but initially focus on particles and branes that are neutral under the graviphoton. Later, we introduce graviphoton charge (KK momentum) and study its consequences.

We begin with the $D$-dimensional action \eqref{eq:Ddimaction} and use the compactification ansatz \eqref{eq:dimredansatz}. Again the Planck mass in the lower dimensional theory is determined by \eqref{eq:Ddconstants}. The kinetic term for the radion is determined by
\begin{equation}
G^{(d)}_{\lambda \lambda} = \frac{(d-1)}{4\kappa_d^2 (d-2)} = M_d^{d-2} \frac{d-1}{4(d-2)}. \label{eq:radionkineticterm}
\end{equation}
The kinetic terms for the other scalars are given by \eqref{eq:Ddconstants}, with no $\lambda$ dependence: the factor of $\e^{-\frac{\lambda}{d-2}}$ from raising an index in $\partial^\mu \phi_j$ compensates the factor from expressing the $D$-dimensional volume factor $\sqrt{-g}$ in $d$-dimensional variables.

Unlike in~\S\ref{eq:dimredWGC}, we keep careful track of $\lambda$-dependent prefactors throughout this section, not just in the action but also in the coupling constants, masses and tensions. This makes the moduli derivatives in~(\ref{eq:repulsion}) easier to evaluate, and provides a useful consistency check. Our final results are easiest to interpret upon setting $\lambda = 0$ by an appropriate rescaling of the constants. Per~\eqref{eq:dimredansatz}, the physical radius of the compact circle is $\e^{-\lambda/2} R$, and likewise $d$-dimensional physical lengths are multiplied by $\e^{\frac{\lambda}{2(d-2)}}$, hence setting $\lambda = 0$ restores the constants to their physical values.

\subsubsection{$P$ preserved}

When $P$ is preserved---i.e., when the brane does not wrap the compact circle---the tension $T_P$ of the $(P-1)$-brane is modified only due to the factor $\e^{\frac{\lambda(x)}{d-2}}$ appearing in the relation between $ds^2$ and $d{\hat s}^2$. This factor rescales $d$-dimensional measurements with respect to the $D$-dimensional measurements. In particular we have, on the worldvolume, $\sqrt{-g} = \sqrt{-{\hat g}}\, \e^{\frac{P \lambda}{2(d-2)}}$, implying
\begin{equation}
T_P^{(d)} = \e^{\frac{P \lambda}{2(d-2)}} T_P^{(D)}.
\end{equation}
The gauge kinetic terms are related by
\begin{equation}
\tau_{ab}^{(d)} = (2 \pi R) \e^{-\frac{P \lambda}{d-2}} \tau_{ab}^{(D)} .
\label{eq:gaugekindimred}
\end{equation}
The factor here comes from the $\lambda$-dependence of $\sqrt{-g}$ in the $D$-dimensional theory multiplied by $P+1$ factors of $\e^{-\frac{\lambda}{d-2}}$ from the raised indices in $F^{\mu_1 \ldots \mu_{P+1}}$. 
Applying~(\ref{eq:repulsion}), the coefficient $\mathcal{P}$ of the self-pressure for this $(P-1)$-brane is
\begin{equation}
\mathcal{P} = (\tau^{(d)})^{ab} q_a q_b - (G^{(d)})^{ij} (\partial_iT_P^{(d)}) ( \partial_j T_P^{(d)}) -(G^{(d)})^{\lambda\lambda} (\partial_\lambda T_P^{(d)})^2 - \frac{P(d-P-2)}{d-2}\frac{\big(T_P^{(d)}\big)^2}{M_d^{d-2}}.  \label{eq:dimreducedselfforce}
\end{equation}
The third term on the right-hand side evaluates to
\begin{equation}
(G^{(d)})^{\lambda\lambda} (\partial_\lambda T_P^{(d)})^2 = \frac{4(d-2)}{M_d^{d-2}(d-1)} \left(\frac{P}{2(d-2)}\right)^2 \big(T_P^{(d)}\big)^2 = \frac{P^2}{(d-1)(d-2)} \frac{\big(T_P^{(d)}\big)^2}{M_d^{d-2}}.
\end{equation}
This combines with the last term in (\ref{eq:dimreducedselfforce}) to give
\begin{equation}
\left[\frac{P^2}{(d-1)(d-2)}  + \frac{P(d-P-2)}{d-2}\right] \frac{\big(T_P^{(d)}\big)^2}{M_d^{d-2}} = \frac{P(D-P-2)}{D-2} \frac{\big(T_P^{(d)}\big)^2}{M_d^{d-2}}.
\end{equation}
Comparing the normalization of $D$-dimensional quantities to $d$-dimensional quantities, we see that the pressure coefficient \eqref{eq:dimreducedselfforce} is precisely the $D$-dimensional pressure coefficient rescaled by an overall factor of $\frac{1}{2\pi R} \e^{\frac{P \lambda}{d-2}}$.

The lesson from this is that the RFC, like the WGC, is exactly preserved under dimensional reduction: the RFC is satisfied for the $P$-form in $d$ dimensions if and only if it is satisfied for the parent $P$-form in $d+1$ dimensions (assuming that the radion remains as a massless modulus). We will now show that the same holds true when $P \rightarrow P -1$ under dimensional reduction.

\subsubsection{$P$ reduced}

When $P$ is reduced---i.e., when the brane wraps the compact circle---the $(P-1)$-brane with tension $T_P^{(D)}$ becomes a $(p-1)$-brane with tension $T_p^{(d)}$, with $p=P-1$. The tensions are related by 
\begin{equation}
T_p^{(d)} = \left(\e^{\frac{\lambda}{d-2}}\right)^{\frac{p}{2}} \left(2 \pi R \e^{-\frac{\lambda}{2}}\right)  T_P^{(D)} = \e^{-\frac{d-p-2}{2(d-2)}\lambda} (2\pi R) T_P^{(D)}.
\label{eq:tensiondimredPred}
\end{equation}
The gauge kinetic terms are related by
\begin{equation}
\tau_{ab}^{(d)} = \frac{1}{2\pi R} \tau_{ab}^{(D)} \e^{\frac{d-p-2}{d-2} \lambda}.
\label{eq:gaugekindimredPred}
\end{equation}
The pressure still has the form of \eqref{eq:dimreducedselfforce} with $P$ replaced by $p$. Again, the term with $\lambda$ derivatives is
\begin{equation}
(G^{(d)})^{\lambda\lambda} (\partial_\lambda T_p^{(d)})^2 = \frac{4(d-2)}{M_d^{d-2}(d-1)} \left(\frac{d-p-2}{2(d-2)}\right)^2 \big(T_p^{(d)}\big)^2 = \frac{(d-p-2)^2}{(d-1)(d-2)} \frac{\big(T_p^{(d)}\big)^2}{M_d^{d-2}}.
\end{equation}
This combines with the last term to give
\begin{equation}
\left[\frac{(d-p-2)^2}{(d-1)(d-2)}  + \frac{p(d-p-2)}{d-2}\right] \frac{\big(T_p^{(d)}\big)^2}{M_d^{d-2}} = \frac{P(D-P-2)}{D-2} \frac{\big(T_p^{(d)}\big)^2}{M_d^{d-2}}.
\end{equation}
Once again, we see that after dividing by $(2 \pi R) \e^{-\frac{d-p-2}{d-2} \lambda}$, the coefficient of the $d$-dimensional pressure \eqref{eq:dimreducedselfforce} for a $(p-1)$-brane matches the coefficient of the $D$-dimensional pressure for a $(P-1)$-brane, and self-repulsiveness is exactly preserved.

One can likewise check that self-repulsiveness is preserved under dimensional reduction for general (non-boost-invariant) branes, see~\cite{BHpaper}.

The preservation of self-repulsiveness under dimensional reduction is an important motivation for the RFC in more than four dimensions. As discussed in appendix \ref{sec:bound}, mutual attraction does not guarantee the existence of a bound state in $D > 4$, hence a theory that violates the RFC in $D>4$ does not necessarily suffer from an infinite tower of self-attractive bound states. This makes the conjecture harder to motivate in higher dimensions. However---since, as we have just seen, self-repulsiveness is preserved under dimensional reduction---given a theory that violates the RFC in $D$ dimensions, reducing on $T^{D-4}$ gives a 4d theory that also violates the RFC. This 4d theory will suffer from an infinite tower of self-attractive bound states by the usual arguments. Thus, in order to avoid such towers, the RFC must be satisfied for all $D \ge 4$.

\subsubsection{Force between general KK modes} \label{subsec:RFCKKmodes}

So far, we have focused on the RFC for a general $P$-form gauge field, ignoring the graviphoton. When the graviphoton is added to the dimensional reduction ansatz for a 1-form or a 2-form in $d+1$ dimensions, the resulting theory in $d$ dimensions will have two 1-form gauge fields: one from the parent theory, and one from the graviphoton. Thus, the graviphoton introduces an additional repulsive force to the theory. On the other hand, the Kaluza-Klein modes that are charged under the graviphoton also receive a contribution to their mass, which increases the attractive gravitational force between them. We will see that these effects precisely cancel, and the self-repulsiveness of each individual KK mode is precisely inherited from the object in the parent theory. However, the force between a particle and its $n$th KK mode becomes attractive in the $R\rightarrow 0$ limit, motivating a tower or sublattice version of the RFC, similar to the WGC.

We begin with the case of 1-form gauge fields in $D = d+1$ dimensions. As in previous subsections, we consider a general number of vector fields $n_v$ and scalar fields $n_s$, with action
\begin{equation}
S_D = \int d^Dx \sqrt{-g} \left[\frac{1}{2 \kappa_D^2} \mathcal{R} - \frac{1}{2} G^{(D)}_{ij} \nabla \phi^i \nabla \phi^j - \frac{1}{2} \tau_{ab}^{(D)} F^a_{\mu\nu} F^{\mu\nu,b}  \right].
\end{equation}
Here, $a, b = 1,...,n_v$ and $i,j=1,...,n_s$. After $S^1$ compactification \eqref{eq:dimredansatz}, the action becomes
\begin{align} 
S_D = \int d^Dx \sqrt{-g} \bigg[\frac{1}{2\kappa_d^2} {\cal R} - \frac{1}{2} G^{(d)}_{\lambda \lambda} \nabla \lambda \nabla \lambda - \frac{1}{2} G^{(d)}_{ij} \nabla \phi^i \nabla \phi^j \nonumber \\
- \frac{1}{2} {\tilde G}^{(d)}_{ab} \nabla \theta^a \nabla \theta^b - \frac{1}{2} \tau^{(d)}_{AB} F^A_{\mu \nu}F^{\mu\nu,B}\bigg].
\end{align}
Here, $A, B \in \{1,...,n_v+1\}$. The number of scalars after compactification is $n'_s = n_s + 1 + n_v$, with one radion $\lambda$ and $n_v$ axions $\theta^a \cong \theta^a + 2\pi$, arising from integrating $A^a$ around the circle direction. The dimensionally reduced kinetic terms in the first line are as previously specified in \eqref{eq:Ddconstants} and \eqref{eq:radionkineticterm}. In the second line, we encounter the axion kinetic matrix
\begin{equation}
\tilde{G}^{(d)}_{ab} = \frac{1}{2 \pi R} \e^\lambda \tau_{ab}^{(D)} = \frac{1}{(2\pi R)^2} \e^{\frac{d-1}{d-2} \lambda} \tau_{ab}^{(d)}.
\end{equation}
We also have the vector kinetic matrix
\begin{equation}
\tau_{AB}^{(d)} = 
\begin{pmatrix}
\tau_{ab}^{(d)} &\quad& \tau^{(d)}_{ab} \big( \frac{\theta^b}{2\pi} \big) \\
 \big( \frac{\theta^a}{2\pi} \big) \tau^{(d)}_{ab}  &\quad& \frac{1}{e_{\text{KK}}^2} + \big( \frac{\theta^a}{2 \pi} \big)  \tau^{(d)}_{ab} \big( \frac{\theta^b}{2 \pi} \big) 
\end{pmatrix} \,,
\end{equation}
with
\begin{equation}
\frac{1}{e_{\text{KK}}^2} = \frac{R^2 M_d^{d-2}}{2}\, \e^{-\frac{d-1}{d-2} \lambda}
\label{eq:KKcoupling}
\end{equation}
and $\tau_{ab}^{(d)}$ as given previously in \eqref{eq:gaugekindimred}, where $P = 1$ in this context. The inverse of the vector kinetic matrix is then
\begin{equation}
\tau_{(d)}^{AB} = 
\begin{pmatrix}
\tau_{(d)}^{ab} + e_{\text{KK}}^2 \big( \frac{\theta^a}{2 \pi} \big) \big( \frac{\theta^b}{2 \pi} \big) & \quad & -e_{\text{KK}}^2 \big(\frac{\theta^b}{2\pi}\big) \\
- e_{\text{KK}}^2 \big(\frac{\theta^a}{2\pi}\big) & \quad & e_{\text{KK}}^2
\end{pmatrix}.
\end{equation}

We would like to compute the force between KK modes. The mass of the $n$th KK mode of a particle of charge $q_a$ and $D$-dimensional mass $m_D$ is given by
\begin{equation}
m_d^2 = \e^{\frac{\lambda}{d-2}} \left(m_D^2 + \e^\lambda \frac{1}{R^2} \left(n - \frac{q_a \theta^a}{2 \pi}  \right)^2 \right) = \e^{\frac{\lambda}{d-2}} \biggl(m_D^2 + \e^\lambda \frac{\tilde{n}^2}{R^2} \biggr) ,
\label{eq:KKmodemass}
\end{equation}
where we have introduced $\tilde n \df n - \frac{q_a \theta^a}{2\pi}$ to declutter our notation below. Now we compute the derivatives of the mass with respect to the $d$-dimensional scalar fields:
\begin{align}
\partial_i m_d &= \frac{m_D}{m_d} \e^\frac{\lambda}{d-2} \partial_i m_D, \nonumber \\
\partial_a m_d &= - \e^{\frac{d-1}{d-2} \lambda} \frac{q_a}{2 \pi} \frac{\tilde n}{m_d R^2},\nonumber \\
\partial_\lambda m_d
&= \frac{d-1}{d-2} \frac{\e^{\frac{\lambda}{d-2}}}{2m_d}  \biggl( \frac{m_D^2}{d-1} + \e^{\lambda} \frac{\tilde{n}^2}{R^2}\biggr),
\end{align}
where $\partial_a \df \frac{\partial}{\partial \theta^a}$.

We can write the requirement \eqref{eq:repulsion} that the force between two KK modes with charges $(q_{1a}, n_1)$ and $(q_{2a}, n_2)$ (which determine their $d$-dimensional masses $m_{d1}$ and $m_{d2}$) be repulsive as
\begin{equation}
\cF_{12}^{(d)} = \cF_{\rm q} - \cF_\phi - \cF_\theta - \cF_\lambda - \cF_{\rm grav} \geq 0,
\label{eq:KKmodeforcetotal}
\end{equation}
where the individual terms are
\begin{align}
\cF_{\rm q} &\df \tau_{(d)}^{AB} q_{1A} q_{2B} = \tau_{(d)}^{ab} q_{1a} q_{2b} + e_{\text{KK}}^2 {\tilde n}_1 {\tilde n}_2 \,, \nonumber \\ 
\cF_\phi &\df G_{(d)}^{ij} \partial_i m_{d1} \partial_j m_{d2} = \e^{\frac{2\lambda}{d-2}} \frac{m_{D1}}{m_{d1}}\frac{m_{D2}}{m_{d2}} G_{(d)}^{ij} \partial_i m_{D1} \partial_j m_{D2}, \nonumber \\
\cF_\theta &\df \tilde{G}_{(d)}^{ab} \partial_a m_{d1} \partial_b m_{d2} = \e^{\frac{d-1}{d-2}\lambda} \tau_{(d)}^{ab} q_{1a} q_{2b} \frac{\tilde{n}_1 \tilde{n}_2}{m_{d1} m_{d2} R^2} , \nonumber  \\
\cF_\lambda &\df G_{(d)}^{\lambda \lambda} \partial_\lambda m_{d1} \partial_\lambda m_{d2} = \e^{\frac{2\lambda}{d-2}} \frac{d-1}{d-2} \frac{\Bigl(\frac{m_{D1}^2}{d-1} +\e^{\lambda} \frac{\tilde{n}_1^2}{R^2} \Bigr)\Bigl(\frac{m_{D2}^2}{d-1}+\e^{\lambda} \frac{\tilde{n}_2^2}{R^2} \Bigr)}{m_{d1}m_{d2} M_d^{d-2}} \,, \nonumber \\
\cF_{\rm grav} &\df \frac{d-3}{d-2} \frac{m_{d1} m_{d2}}{M_d^{d-2}}.  \label{eq:KKmodeforces}
\end{align}
After some simplification, we obtain
\begin{align}
\cF_{12}^{(d)} &= \frac{m_{d1} m_{d2} - \e^{\frac{d-1}{d-2}\lambda} \frac{\tilde{n}_1 \tilde{n}_2}{R^2}}{m_{d1} m_{d2}} \tau_{(d)}^{a b} q_{1 a} q_{2 b}
-\e^{\frac{d-1}{d-2}\lambda} \frac{\Bigl[\sqrt{\frac{m_{d2}}{m_{d1}}} \frac{\tilde{n}_1}{R} - \sqrt{\frac{m_{d1}}{m_{d2}}} \frac{\tilde{n}_2}{R}\Bigr]^2}{M_d^{d-2}}
\nonumber \\
&\noeq -\e^{\frac{2 \lambda}{d-2}} \frac{m_{D1} m_{D2}}{m_{d1} m_{d2}} \biggl[G_{(d)}^{i j} \partial_i m_{D1} \partial_j m_{D2} 
+\frac{d-2}{d-1} \frac{m_{D1}m_{D2}}{M_d^{d-2}} \biggr] \,. \label{eqn:KKtotalforce}
\end{align}
This expression simplifies considerably in the self-force case $m_{D1} = m_{D2}$, $q_1 = q_2$, $n_1 = n_2$, where the second term vanishes and the first term simplifies:
\begin{align}
\cF_{11}^{(d)} &= \e^{\frac{2 \lambda}{d-2}} \frac{m_D^2}{m_d^2} \biggl[\e^{-\frac{\lambda}{d-2}} \tau_{(d)}^{a b} q_{a} q_{b} 
- G_{(d)}^{i j} \partial_i m_D \partial_j m_D - \frac{d-2}{d-1} \frac{m_D^2}{M_d^{d-2}} \biggr] \,.
\end{align}
Using \eqref{eq:gaugekindimred}, \eqref{eq:Ddconstants}, and the fact that $d = D - 1$, we see that this is precisely the self-force coefficient in $D$ dimensions \eqref{eq:repulsion} multiplied by the factor $\frac{\e^{2\lambda/(d-2)}}{2\pi R} \frac{m_D^2}{m_d^2}$. 

Thus, provided the particle was self-repulsive in $D$-dimensions, all of its KK modes will be self-repulsive. However, as we discuss in \S\ref{sec:STRONG} below, due to the second term in \eqref{eqn:KKtotalforce} two different KK modes of the same particle are not necessarily mutually repulsive. This will motivate us to consider strong forms of the RFC.

\subsubsection{Force between KK and winding modes}

We now reexamine the case where the $D$-dimensional theory has $N$ two-form gauge fields and associated charged strings. We begin with the $D$-dimensional action,
\begin{equation}
S_D = \int d^Dx \sqrt{-g} \left[\frac{1}{2 \kappa_D^2} \mathcal{R} - \frac{1}{2} G_{ij} \nabla \phi^i \nabla \phi^j - \frac{1}{2 \cdot 3!} \tau_{ab} (F_{\mu_1 \mu_2  \mu_3}^a  F^{b,\mu_1 \mu_2 \mu_3})  \right] .
\end{equation}
Upon reduction to $d$ dimensions, we obtain $N+1$ one-form gauge fields: one for each two-form in $D$ dimensions, as well as the graviphoton. The kinetic matrix is
\begin{equation}
\tau_{AB}^{(d)} = \left( 
\begin{array}{cc}
\tau_{ab}^{(d)} & 0 \\
0  & \frac{1}{e_{\text{KK}}^2}
\end{array}
\right),
\end{equation}
with $e_{\text{KK}}^2$ given by \eqref{eq:KKcoupling} and $\tau_{ab}^{(d)}$ given by \eqref{eq:gaugekindimredPred} with $p = 1$. This is simpler than before, as there are no axions to induce kinetic mixing with the graviphoton.

Wound strings and the KK modes of the graviton give rise to the spectra,
\begin{align}
m_{\text{str}} &= \e^{-\frac{d-3}{2(d-2)}\lambda} (2 \pi R) T \,, & m_n &= \e^{\frac{d-1}{2(d-2)}\lambda} \frac{|n|}{R}\,,
\end{align}
respectively (see~\eqref{eq:tensiondimredPred},~\eqref{eq:KKmodemass}) where $n$ is any integer. Because of the absence of kinetic mixing, the mutual force between a wound string and a KK graviton has no gauge contribution. Likewise, because the KK graviton mass is independent of the $D$-dimensional moduli, the scalar contribution to the mutual force is mediated solely by the radion. Applying~\eqref{eq:repulsion} and~\eqref{eq:radionkineticterm}, we obtain:
\begin{equation}
\mathcal{F}_{\text{str},n} = -\frac{4 (d-2)}{(d-1) M_d^{d-2}} \biggl[\frac{-(d-3) m_{\text{str}}}{2(d-2)}\biggr]\biggl[ \frac{(d-1)m_n}{2(d-2)}\biggr] - \frac{d-3}{d-2} \frac{m_{\text{str}} m_n}{M_d^{d-2}} = 0\,.
\end{equation}
Thus, the mutual force vanishes due to a cancellation between the radion and graviton contributions. In particular, the radion force is repulsive because the wound string and KK gravitons couple to the radion with opposite sign: KK modes are light at large $R$ whereas the wound string is heavy, and vice versa at small $R$.

In many cases, it is possible to give the wound string nonzero KK momentum around the compact circle. Heuristically, this can be thought of as a ``bound state'' of a KK graviton with the wound string. The vanishing of the long range force between the two constituents suggests the mass formula
\begin{equation} \label{eqn:KKstring}
m_{\text{str}}^{(n)} = \e^{-\frac{d-3}{2(d-2)}\lambda} (2 \pi R) T + \e^{\frac{d-1}{2(d-2)}\lambda} \frac{|n|}{R} \,,
\end{equation}
such that the binding energy vanishes. Noting that this formula correctly describes (part of) the spectrum of tree-level string theory on a compact circle, we analyze its consequences without claiming it to be completely general.

The coefficient of the mutual force between two such wound strings is
\begin{equation}
\mathcal{F}_{12}^{(d)} = \tau^{ab}_{(d)} q_{1a} q_{2b} + e_{\text{KK}}^2 n_1 n_2 - G^{(d)}_{ij} \partial_i m_1 \partial_j m_2 
- G^{\lambda \lambda} \partial_\lambda m_1 \partial_\lambda m_2 - \frac{d-3}{d-2} \frac{m_1 m_2}{M_d^{d-2}} \,.
\end{equation}
Applying~\eqref{eqn:KKstring} as well as~\eqref{eq:Ddconstants}, \eqref{eq:gaugekindimredPred}, \eqref{eq:radionkineticterm}, and~\eqref{eq:KKcoupling}, we obtain
\begin{equation}
\mathcal{F}_{12}^{(d)} = (2\pi R) \e^{-\frac{d-3}{d-2}\lambda} \biggl[\tau_{(D)}^{a b} q_{1a} q_{2b} - G_{(D)}^{i j} \partial_i T_1 \partial_j T_2 - \frac{2(D-4)}{D-2} \frac{T_1 T_2}{M_{D}^{D-2}} \biggr] - e_{\text{KK}}^2 (|n_1 n_2| - n_1 n_2) \,,
\end{equation}
after a straightforward computation. We recognize the term in brackets as the coefficient of the mutual pressure (\eqref{eq:repulsion} with $P=2$) between the strings in $D$ dimensions.

This result is far simpler than~\eqref{eqn:KKtotalforce}! String winding modes with KK charge of the same sign will be mutually repulsive if and only if the strings are mutually repulsive in the parent, $D$-dimensional theory, whereas wound strings with opposite-sign KK charges experience an additional attractive force. This result matches our expectations from type II superstring theory, where the momentum and winding modes are mutually BPS. In less supersymmetric contexts, the formula~\eqref{eqn:KKstring} could be modified, and the behavior of modes with both momentum and winding correspondingly altered (if such modes continue to exist). However, it is worth noting that parametrically large extremal black holes can carry both momentum and winding charge, and always obey the formula~\eqref{eqn:KKstring}, see~\cite{Heidenreich:2015nta}.

\subsection{Gauge-scalar-gravity unification}

We now examine the self-force implications of emergence from an ultraviolet cutoff~\cite{Harlow:2015lma, Heidenreich:2017sim}. By an argument similar to that of~\S\ref{ssec:gaugegrav}, we find that this implies the existence of at least one particle for which gauge repulsion is not parametrically less than the scalar and gravitational attractions, so that the RFC cannot be parametrically violated.

As with the gauge and gravitational forces, we define the strong-coupling scale $\Lphi$ of the scalar field to be the scale at which the 1PI corrections to the scalar propagator from particles lighter than $\Lphi$ rivals the tree-level contribution, and the parameter~\cite{Heidenreich:2018kpg}
\begin{equation}
\lambda_{\phi}(\Lambda)\df   \Lambda^{D-4} G^{\phi \phi}(\phi) \sum_{i | m_i < \Lambda} \biggl(\frac{\partial m_i}{\partial \phi}\biggr)^2
\end{equation}
is equal to 1, i.e., $\lambda_{\phi}(\Lphi) \df 1$.\footnote{In the notation of \cite{Heidenreich:2018kpg}, $G^{\phi \phi}(\phi)$ was denoted $1/K(\phi)$.}

Gauge-scalar unification in the sense of~\cite{Heidenreich:2017sim, Heidenreich:2018kpg} is the assumption that $\Lphi \sim \Lgauge$, where $\Lgauge$ was defined in~\S\ref{ssec:gaugegrav}, see (\ref{eq:Landaupole}). This immediately implies
\begin{align}
e^2 \langle  q^2 \rangle_{\Lgauge} \sim G^{\phi \phi}(\phi) \biggl\langle  \biggl(\frac{\partial m_i}{\partial \phi}\biggr)^2 \biggr\rangle_{\Lgauge} ,
\label{eqn:agreement}
\end{align}
where the average is taken over particles lighter than $\Lgauge$.
We can interpret the left-hand side of this equation as the average gauge force between the light particles, and the right-hand side as the average scalar force between the light particles. Thus, at least one such particle must have a gauge self-force that is not parametrically smaller than its scalar self-force.

As previously argued, gauge-gravity unification ($\Lgauge \sim \Lgrav$) implies~(\ref{eqn:gaugegrav}). In combination with~(\ref{eqn:agreement}), this implies
\begin{equation}
e^2 \langle  q^2 \rangle_{\Lgauge} \sim G^{\phi \phi}(\phi) \biggl\langle  \biggl(\frac{\partial m_i}{\partial \phi}\biggr)^2 \biggr\rangle_{\Lgauge} + \frac{D-3}{D-2} \frac{\Lgauge^2}{M_D^{D-2}} ,
\end{equation}
up to order-one factors, since the left-hand side is parametrically of the same order as \emph{each term} on the right-hand side. Since by definition $m_i \le \Lgauge$ for the light particles, this implies that the average light particle is self-repulsive, and in particular at least one particle in the spectrum must be self-repulsive, up to order-one factors in either case. Thus, the RFC cannot be parametrically violated in emergent theories of the kind we have described.

Conversely, suppose that the light spectrum is dominated by a tower of charged, self-repulsive particles. This implies that
\begin{equation}
\lambda_\phi(\Lambda) \sim \Lambda^{D-4} G^{\phi \phi}(\phi) \sum_{n|m_n< \Lambda} \left( \frac{\partial m_n}{\partial \phi} \right)^2  \lesssim {\Lambda^{D-4}} e^2  \sum_{n|m_n< \Lambda} q_n ^2 \sim \lambda_{\rm gauge}(\Lambda) \,,
\end{equation}
and so $\Lgauge \lesssim \Lphi$. By a similar argument, the same assumptions lead to $\Lgauge \lesssim \Lgrav$, as in~\cite{Heidenreich:2017sim}.

\section{The RFC and black holes}

The WGC is defined with respect to large, extremal black holes. The RFC, on the other hand, is defined by long range forces, and makes no reference to black holes. It is worth asking, therefore, whether any connection between the RFC and extremal black holes persists in the presence of massless scalars.

As previously discussed, extremal Reissner-Nordstr\"om black holes holes have zero self-force while subextremal black holes are self-attractive. In this section, we will show that the same is true in (two-derivative) Einstein-Maxwell-dilaton gravity. In \cite{BHpaper}, this is shown to generalize to an arbitrary two-derivative action with gauge fields and moduli, as well as to higher $p$-forms. Once again, we stress that the vanishing of the self-force between large extremal black holes does not imply that the WGC and RFC are equivalent, see figure \ref{fig:phase}. Nonetheless, it does suggest that the conjectures remain closely related, even in the presence of massless scalars.

Consider Einstein-Maxwell-dilaton gravity, with action given by
\begin{equation}
S = \frac{1}{2 \kappa_D^2} \int d^Dx \sqrt{-g} \left(R - \frac{1}{2}(\nabla \phi)^2 \right) - \frac{1}{4 e^2} \int d^Dx\sqrt{-g} \e^{-\alpha \phi} F_{\mu\nu}F^{\mu\nu}.
\end{equation}
The black hole extremality bound is \cite{horowitz:1991cd,myers:1986un,gibbons:1987ps, Heidenreich:2015nta}:
\begin{equation}
  \gamma e^2 q^2 M_D^{D - 2} \leq m^2 \;,
\qquad \text{where} \qquad
  \gamma \df \left[ \frac{\alpha^2}{2} + \frac{D - 3}{D - 2} \right]^{- 1}
  \, . \label{eqn:gamma}
\end{equation}
In the conventions of \cite{Heidenreich:2015nta}, the black hole geometry has the form:
\begin{align}
  g_{tt} &= - \left[ 1 - \left( \frac{r_+}{r} \right)^{D - 3} - \left[ \frac{2
  (D - 3) \gamma}{D - 2} - 1 \right]  \left( \frac{r_-}{r} \right)^{D - 3} 
  \right]  + \ldots \,, \nonumber \\
  A_t &= \frac{e}{\kappa_D}  \sqrt{\gamma}  \frac{(r_+ r_-)^{\frac{D -
  3}{2}}}{r^{D - 3}} \,,
  \qquad \quad \phi = - \alpha \gamma \left( \frac{r_-}{r} \right)^{D - 3} + \ldots \,,   
\end{align}
to leading order in large $r$.
The mass, charge, and scalar charge of the black hole can be read off from this asymptotic behavior,
\begin{align}
  m &= \frac{V_{D - 2}}{2 \kappa_D^2} [(D - 2) (r_+^{D - 3} - r_-^{D - 3}) +
  2 (D - 3) \gamma r_-^{D - 3}] \,, \nonumber \\
  q &= \frac{(D - 3) V_{D - 2}}{e \kappa_D}  \sqrt{\gamma}  (r_+
  r_-)^{\frac{D - 3}{2}} \,,
  \qquad \quad [\partial_{\phi} m]_{\mathrm{eff}} = \frac{(D - 3) V_{D - 2}}{2
  \kappa_D^2} \alpha \gamma r_-^{D - 3} \,,   \label{eqn:BHconstants}
\end{align}
where we used $G_{\phi \phi}=1/(2 \kappa_D^2)$
for the conventionally-normalized dilaton. Here we denote the scalar charge as $[\partial_{\phi} m]_{\mathrm{eff}}$ in accordance with the general results of~\S\ref{subsec:forces}.

The mass and charge formulae agree with those in \cite{Heidenreich:2015nta}. The formula for the scalar charge can be explained as follows.
The black hole entropy is proportional to the horizon area,
which is
\begin{equation}
  A = V_{D - 2} (r^2_+ f_-^{\gamma_{\perp}} (r_+))^{\frac{D - 2}{2}} = V_{D -
  2}  \left( r^{D - 3}_+  \left[ 1 - \left( \frac{r_-}{r_+} \right)^{D - 3}
  \right]^{\frac{\alpha^2 \gamma}{2}} \right)^{\frac{D - 2}{D - 3}} \; .
\end{equation}
As the black hole moves adiabatically through a scalar gradient, the gauge
coupling $e$ can change, but the quantized charge $q$ must remain invariant.
Moreover, we expect that the internal state of the black hole is unaffected,
so the black hole entropy, and hence the horizon area, should also remain
unchanged. Holding $A, q$ fixed and allowing $r_{\pm}$ and $e$ to vary, we
find
\begin{align}
  \frac{\delta r^{D - 3}_+}{r_+^{D - 3}} + \frac{\alpha^2 \gamma}{2}  \frac{-
  \delta \frac{r_-^{D - 3}}{r_+^{D - 3}}}{1 - \frac{r_-^{D - 3}}{r_+^{D - 3}}}
  &= 0 \,, &  \frac{\delta r^{D - 3}_+}{r_+^{D - 3}} + \frac{\delta
  r^{D - 3}_-}{r_-^{D - 3}} &= \frac{\delta e^2}{e^2} \; .
\end{align}
After some manipulation, we obtain,
\begin{align}
  \delta m &= \frac{V_{D - 2}}{2 \kappa_D^2} [(D - 2) (\delta r_+^{D - 3} -
  \delta r_-^{D - 3}) + 2 (D - 3) \gamma \delta r_-^{D - 3}] \nonumber \\
  &= \frac{\gamma (D - 3) V_{D - 2}}{2 \kappa_D^2} r_-^{D - 3}  \frac{\delta
  e^2}{e^2} = \frac{\alpha \gamma (D - 3) V_{D - 2}}{2 \kappa_D^2} r_-^{D - 3}
  \delta \phi \,,
\end{align}
where in the last step we use $e^2 (\phi) \propto e^{\alpha \phi}$. We read
off
\begin{equation}
  \left. \frac{\partial m}{\partial \phi} \right|_{q, S} = \frac{(D - 3) V_{D
  - 2}}{2 \kappa_D^2} \alpha \gamma r_-^{D - 3}
\end{equation}
in agreement with (\ref{eqn:BHconstants}). Thus, we can interpret
$[\partial_{\phi} m]_{\mathrm{eff}}$ as $\partial_{\phi} m$ with the black hole
charge and entropy held fixed, in agreement with the physical arguments given
above.

The mass $m$, charge $q$, and scalar charge $[\partial_{\phi} m]_{\mathrm{eff}}$ given in (\ref{eqn:BHconstants}) are related as follows:
\begin{equation}
  \frac{e^2 q^2}{2 \kappa_D^2} = [\partial_{\phi} m]_{\mathrm{eff}}^2 + \frac{2
  (D - 3)}{D - 2}  \left( m - \frac{[\partial_{\phi} m]_{\mathrm{eff}}}{\alpha}
  \right)  \frac{[\partial_{\phi} m]_{\mathrm{eff}}}{\alpha} \;,
  \label{eqn:BHeffdil}
\end{equation}
whereas the black hole is self-attractive if
\begin{equation}
  \frac{e^2 q^2}{2 \kappa_D^2} < [\partial_{\phi} m]_{\mathrm{eff}}^2 + \frac{D
  - 3}{2 (D - 2) } m^2 \; . \label{eqn:BHselfattract}
\end{equation}
Using (\ref{eqn:BHeffdil}) to eliminate $q^2$ from (\ref{eqn:BHselfattract}),
we find that the BH is self-attractive if
\begin{equation}
  0 < \frac{D - 3}{2 (D - 2) }  \left( m - \frac{2}{\alpha}  [\partial_{\phi}
  m]_{\mathrm{eff}} \right)^2 \;, \qquad \text{or equivalently } \; m \neq
  \frac{2}{\alpha}  [\partial_{\phi} m]_{\mathrm{eff}} \; .
\end{equation}
Putting $[\partial_{\phi} m]_{\mathrm{eff}} = \frac{\alpha}{2} m$ into
(\ref{eqn:BHeffdil}) we obtain
$\gamma e^2 q^2 = \kappa_D^2 m^2$,
so the black hole is extremal. Thus, in this theory
sub-extremal black holes are self-attractive and extremal black holes
have no long-range force between them.

It is interesting to ask whether we can violate cosmic censorship by
introducing a particle with charge-to-mass ratio $q / m > | q / m
|_{\mathrm{ext}}$, but with $\partial_{\phi} m$ large enough (and of the right
sign) so that the particle is nonetheless attracted to an extremal black
hole. In fact, this is not possible, because a minimum energy
$E \ge q \Phi_H$ is required for the particle to cross the horizon (see, e.g.,~\cite{BHpaper}), where
\begin{equation}
  \Phi_H = A_t (r_+) - A_t (\infty) = \frac{e}{\kappa_D}  \sqrt{\gamma} 
  \left( \frac{r_-}{r_+} \right)^{\frac{D - 3}{2}}
\end{equation}
is the electrostatic potential at the horizon and $E$ is the total energy of
the particle (including its rest mass). This means that
\begin{equation}
  \gamma e^2 q^2 M_D^{D - 2} \le \frac{r_+^{D - 3}}{r_-^{D - 3}} E^2 \,,
\end{equation}
so in particular, an extremal black hole ($r_+ = r_-$) can absorb only
subextremal particles, or superextremal particles with enough kinetic energy to make their charge-to-energy ratio subextremal.

An intriguing consequence of this is that, if the situation described above occurs in four dimensions, the superextremal particle and the black hole can form a stable, non-rotating, superextremal bound state.

\subsection{Finite-size effects and a related conjecture}

\begin{figure}
\centering
\includegraphics[width=7cm]{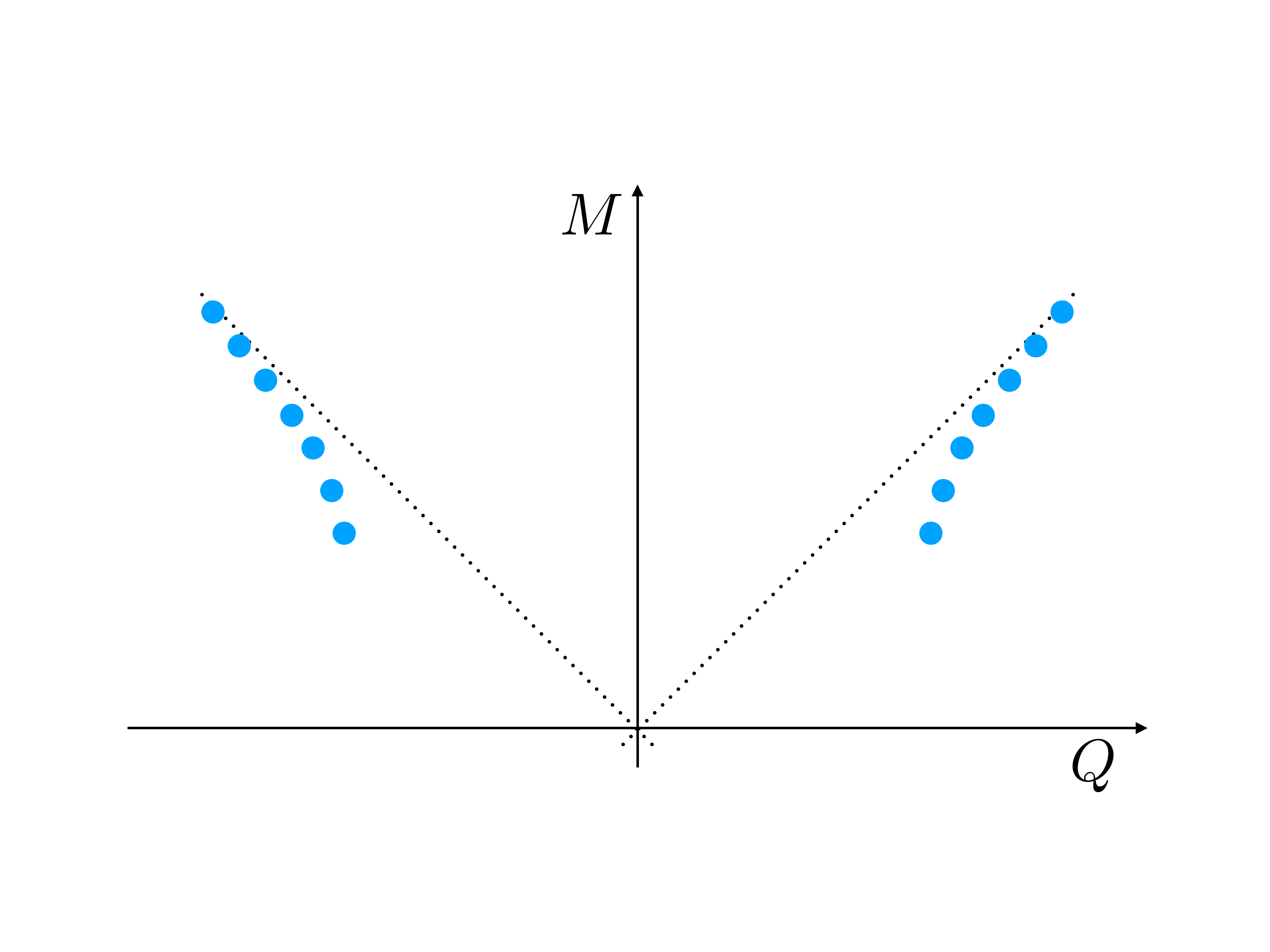}
\includegraphics[width=7cm]{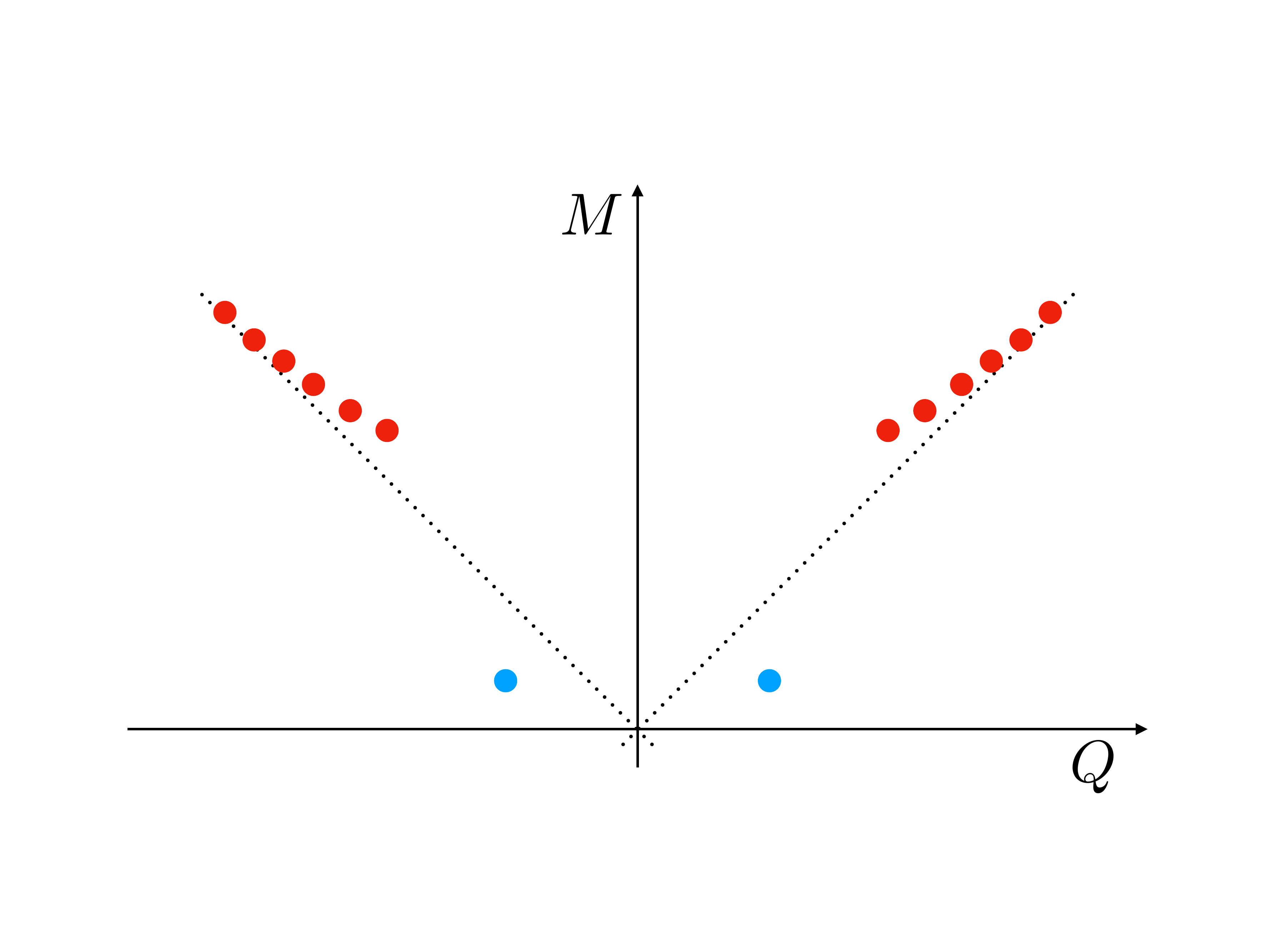}
\caption{Higher-dimension operators introduce corrections to the extremality bound for finite-sized black holes. In principle, these may increase the charge-to-mass ratio $Q/M$ (left) or decrease it (right). The WGC is necessarily satisfied in the former case, whereas it will be violated unless there exist light superextremal particles (blue) in the latter case.}
\label{fig:BH spectrum}
\end{figure}

The WGC is closely related to finite-size effects for black holes. These effects can be generated by massive particles, loops of massless particles, and bare higher-derivative couplings
 in the Lagrangian, and can in principle either increase or decrease the maximum possible charge-to-mass ratio $\vec{Q}/M$ for finite-sized black holes. In the former case, the WGC is necessarily satisfied, either by maximally charged black holes or by a stable decay product thereof.\footnote{Note that, as stated in section~\S\ref{subsec:WGC}, we define ``extremal'' to mean an object whose charge-to-mass ratio is that of a maximally charged, parametrically large black hole. Thus, a maximally charged, finite size black hole is not necessarily extremal. It can be either subextremal or (super)extremal, depending on finite-size effects, where as usual, we define ``superextremal'' to include the exactly extremal case.} In the latter case, however, the WGC may in principle be violated if there are no light, superextremal particles, since no finite-sized black holes satisfy the WGC bound. This is depicted in figure \ref{fig:BH spectrum}.

The leading-order finite-size corrections can be encoded in four-derivative terms in the action of the schematic form $F^4$, $F^2 R$, and $R^2$. The precise linear combination of these terms that appears in the charge-to-mass ratio of a maximally-charged black hole in theories without massless scalars was worked out in \cite{kats:2006xp}. In \cite{Cheung:2018cwt,Bellazzini:2019xts,Mirbabayi:2019iae,asymptotic}, it was argued that the sign of this linear combination is fixed so that maximally-charged black holes of finite size are always superextremal, implying the WGC.

In the case of the RFC, a similar statement is true: finite-size effects modify the self-repulsiveness of maximally-charged black holes. In the absence of massless scalars, the self-force depends only on the conserved charge and mass, and so these effects lead to self-repulsive black holes precisely when they lead to superextremal black holes. With massless scalars, neither the corrections to extremality nor to the self-force have been studied in detail to date. It would be interesting to explore the linear combinations of four-derivative operators that correct the self-force and the charge-to-mass ratio in the presence of scalars and see whether they are related and if either or both have a definite sign.

It is natural to expect---by analogy with~\cite{kats:2006xp}---that whenever these corrections are nonzero in an actual quantum gravity, they cause maximally-charged black holes to be self-repulsive (see, e.g.,~\cite{Horne:1992bi}). If instead maximally-charged black holes were self-attractive, then two identical such black holes would attract each other. This would induce some sort of gravitational collapse, the probable outcome of which would be a single black hole of twice the charge.\footnote{Alternately, the final configuration could be a stable, non-rotating, multicenter solution. In pure gravity this would be in tension with various black hole uniqueness theorems. These theorems may or may not generalize in some form to theories with moduli and higher-derivative corrections.} Energy conservation implies that this black hole would have a larger charge-to-mass ratio than the original (less massive) one, and therefore this scenario is only possible if finite-sized black holes are \emph{subextremal}. Thus, there is some relation between this conjecture and the analogous one~\cite{kats:2006xp} about finite-size corrections to the charge-to-mass ratio of maximally-charged black holes.

\section{Strong forms of the WGC and RFC}\label{sec:STRONG}

\subsection{Review of strong forms of the WGC}

``Strong forms" of the WGC have been discussed at length, motivated in large part by their potential ability to constrain models of axion inflation \cite{Arkanihamed:2006dz, rudelius:2014wla, Delafuente:2014aca, rudelius:2015xta, Brown:2015iha, Heidenreich:2015wga, Montero:2015ofa,Ibanez:2015fcv,Bachlechner:2015qja,banks:2003sx,Brown:2015lia,junghans:2015hba,Hebecker:2015zss,Hebecker:2015rya, Conlon:2016aea,Hebecker:2017uix,Blumenhagen:2017cxt,Hebecker:2018fln, Hebecker:2018yxs, Buratti:2018xjt}. Although a number of strong forms have been falsified \cite{Heidenreich:2016aqi}, there is a growing body of evidence in favor of a pair of closely-related strong forms: the Sublattice WGC (sLWGC)~\cite{Heidenreich:2016aqi} and the Tower WGC (TWGC)~\cite{Andriolo:2018lvp}.

The TWGC is the strictly weaker of the two: essentially, it requires not just one superextremal particle, but rather an infinite tower of them. The conjecture can be satisfied by unstable resonances, but (unlike the mild WGC) not by multiparticle states. At weak coupling, the resonances will be narrow and their existence and charge-to-mass ratio can be sharply defined. Away from weak coupling the precise meaning of the conjecture---and of the sLWGC, for the same reasons---is uncertain.

\begin{figure}
\centering
\includegraphics[width=6cm]{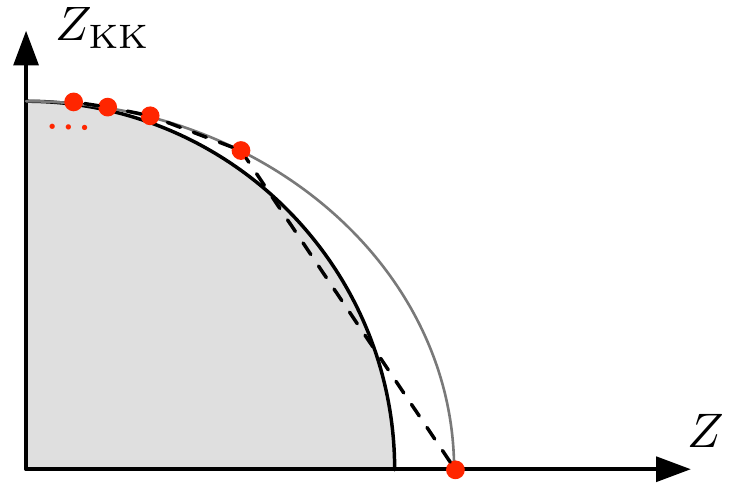}
\includegraphics[width=6cm]{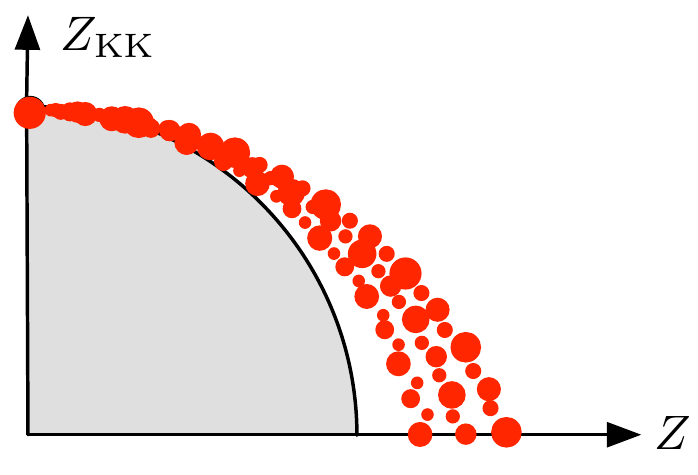}
\caption{The CHC can be violated by KK reduction on a small circle, even if the WGC is satisfied in the parent theory. In particular, in the small $R$ limit, the KK modes of a superextremal particle collect near the ``poles'' of the black hole region, violating the CHC (left). This problem can be avoided by postulating an infinite tower of superextremal particles of increasing charge, whose KK modes densely fill in the directions in charge space.}
\label{fig:CHCviolation}
\end{figure}

It is useful to make a somewhat more precise statement.
One motivation for the existence of such a tower of superextremal particles is the observation that its absence in $D=d+1$ dimensions generally leads to a violation of the ordinary WGC in $d$ dimensions after Kaluza-Klein reduction on a circle~\cite{Heidenreich:2015nta}.\footnote{Additional motivations were given in~\cite{Andriolo:2018lvp}.} In \S\ref{eq:dimredWGC}, we argued that---accounting for graviphoton charge and radion couplings---all the KK modes of a superextremal particle are superextremal. 
 However, although individual KK modes may be superextremal, this is not sufficient to ensure that the CHC will be satisfied in the $R \rightarrow 0$ limit after Kaluza-Klein reduction. In this limit, almost all of the KK modes of any finite set of charged particles accumulate near the ``poles'' of the black hole region, violating the CHC as illustrated in figure \ref{fig:CHCviolation} (left). This problem can be avoided by mandating an infinite tower of superextremal particles in $D$ dimensions, as shown in figure \ref{fig:CHCviolation} (right).

However, even demanding an infinite tower of superextremal particles of increasing mass does not guarantee consistency under dimensional reduction. Imagine a $U(1)$ theory in $D$ dimensions in which lattice sites of charge $3 n$, for $n \in \mathbb{Z}$, are completely devoid of superextremal particles. Then, upon $S^1$ compactification, consider the charge $(3,1)$ direction in charge space, where the $3$ represents the charge under the original $U(1)$ and the $1$ is the Kaluza-Klein charge. Since there were no superextremal particles of charge $3 n$ to begin with, there will not be any superextremal KK modes in this direction in charge space, and the CHC can be violated at small $R$.

To ensure that the WGC is satisfied after dimensional reduction, it is sufficient to exclude this possibility, motivating the following definition of the TWGC:
\begin{namedconjecture}[The Tower Weak Gravity Conjecture (TWGC)]
For every site in the charge lattice, $\vec{q} \in \Gamma$, there exists a positive integer $n$ such that there is a superextremal particle of charge $n \vec{q}$.
\end{namedconjecture}
\noindent
Since there is a superextremal resonance in every rational direction in the charge lattice, the final state from the decay of this resonance (or the resonance itself, if it is stable) is a superextremal multiparticle state, and the WGC is satisfied in $D$ dimensions. Moreover, this conjecture is necessary and sufficient to ensure that there is a superextremal KK mode in every rational direction of the charge lattice after compactification on a circle, and so the WGC is satisfied in $d= D-1$ dimensions. These KK modes likewise ensure that the TWGC itself is satisfied in $d=D-1$ dimensions,\footnote{Here we ignore quantum corrections in the $d$-dimensional theory. This is particular important upon compactification to four (or fewer) dimensions, as discussed below.} and so the WGC remains true after compactification on a torus, etc.
This definition also ensures an infinite tower of particles in each direction in charge space, consistent with the general idea of the conjecture given above.

The sLWGC is strictly stronger than the TWGC: it requires a (full-dimensional) sublattice of the charge lattice for such that there is a superextremal particle at each site. In other words, the integer $n$ appearing in the definition of the TWGC can be taken to be universal, i.e., independent of $\vec{q}$:
\begin{namedconjecture}[The Sublattice Weak Gravity Conjecture (sLWGC)]
There exists a positive integer $n$ such that for any site in the charge lattice, $\vec{q} \in \Gamma$, there is a superextremal particle of charge $n \vec{q}$.
\end{namedconjecture}
\noindent
Implicit in this conjecture is the idea that $n$ is not parametrically large, but no sharp limits on it are known.

Much of the evidence in favor of the WGC can actually be used in support of these strong forms of the conjecture. The modular invariance argument of \S\ref{ssec:modular} implies a sublattice full of superextremal states, and many examples in string theory satisfy the sLWGC. The emergence argument of \S\ref{ssec:gaugegrav} similarly implies the existence of a tower of states that satisfy the WGC bound on average (up to order-one factors), which is closely related to the TWGC. Calabi-Yau three-fold compactifications of type IIA string theory \cite{Grimm:2018ohb} and F-theory \cite{Lee:2018spm,Lee:2018urn} have been argued to support an infinite tower of superextremal states, and infrared consistency has been used to argue that quantum gravity theories must have a tower of superextremal particles in the event that all charged particles are scalar fields \cite{Andriolo:2018lvp}.

Unlike the ordinary WGC, the TWGC and sLWGC are both preserved under dimensional reduction at tree-level. In four dimensions, however, there is an important subtlety~\cite{Heidenreich:2016aqi, conifolds}: massless charged particles logarithmically renormalize the gauge coupling to zero in the deep infrared. Technically, this represents a counterexample to the TWGC and sLWGC because the gauge coupling vanishes in the deep infrared, yet there is no infinite tower of massless particles.\footnote{The mild WGC is satisfied, since by assumption there is a massless charged particle. More generally, the log running makes very light charged particles exponentially superextremal.} However, this is a fairly benign counterexample, and such theories typically satisfy some sort of renormalized version of the T/sLWGC, in which we allow the gauge coupling $e=e(\Lambda)$ appearing in the WGC bound to depend on the energy scale (see, e.g.,~\cite{Heidenreich:2017sim} for a brief discussion).

A more interesting potential counterexample to these conjectures in a 4d F-theory compactification appeared in \cite{Lee:2019tst}: although the full spectrum of the theory in question could not be computed, the sector considered contained an infinite tower of superextremal particles that did not satisfy the precise stipulations of the T/sLWGC as we have defined them above. While it is possible that the theory might satisfy the T/sLWGC once all sectors are included, it is worth noting that a counterexample to these conjectures in 4d would not be too surprising, since the T/sLWGC in $D$ dimensions are intimately related to the WGC in $d=D-1$ dimensions, and it is not clear that the WGC should hold (or, indeed, what the conjecture is, precisely) for $d \leq 3$.

\subsection{Strong forms of the RFC}\label{ssec:RFCSTRONG}

Dimensional reduction of the RFC leads to a similar conclusion as for the WGC: compactification on a small circle can lead to a violation of the conjecture, requiring a ``strong form." To see this, consider the force between the 0th and $n$th KK modes of a particle charged under a $1$-form after compactification from $D$ to $d=D-1$ dimensions, setting $\theta^a = 0$ for simplicity. From~\eqref{eqn:KKtotalforce}, we obtain (setting $\lambda = 0$)
\begin{align}
\cF_{0n} &= \tau_{(d)}^{a b} q_{a} q_{b}
-\frac{m_D}{\sqrt{m_D^2+\frac{n^2}{R^2}}} \biggl[G_{(d)}^{i j} \partial_i m_{D} \partial_j m_{D} 
+\frac{d-2}{d-1} \frac{m_{D}^2}{M_d^{d-2}} +\frac{n^2}{R^2 M_d^{d-2}} \biggr] \,.
\end{align}
Now consider the $R\rightarrow 0 $ limit.\footnote{Note that in terms of $D$-dimensional quantities, $1/M_d^{d-2}, (\tau^{(d)})^{ab},$ and $(G^{(d)})^{ij}$ all scale as $1/R$, so whether we hold $D$- or $d$-dimensional kinetic terms fixed only affects the overall scaling with $R$ and not the form of this inequality.} The inequality $\cF_{0n} \geq 0$ becomes:
\begin{equation}
\tau_{(d)}^{ab} q_a q_b \geq \frac{n}{R}\cdot\frac{m_D}{M_d^{d-2}} + O(R).
\label{eq:R0KK}
\end{equation}
For any nonzero $m_D$, the inequality is violated for sufficiently small $R$. The precise value of $R$ at which the force becomes attractive depends on $n$ and the mass, charge, and scalar charge, but in any case one can check that it is no smaller than
\begin{equation}
R_{\text{crit}} \df \frac{m_D}{\tau^{a b}_{(d)} q_a q_b M_d^{d-2}} \,.
\end{equation}
For $R<R_{\text{crit}}$, the KK zero mode attracts all the other KK modes.
Likewise, for any two modes KK charges of opposite sign ($n_1>0$ and $n_2 < 0$ or vice versa), the mutual force~\eqref{eqn:KKtotalforce} is bounded by:
\begin{align}
\cF_{12}^{(d)} \le 2 \tau_{(d)}^{a b} q_{a} q_{b} - 4 \frac{|n_1| |n_2|}{R^2 M_d^{d-2}} \,,
\end{align}
and so the force is attractive for any $R^2 < R_{\text{crit}}^{\prime 2} \df \frac{2}{\tau_{(d)}^{a b} q_{a} q_{b}M_d^{d-2}}$.

As a result, the RFC can be violated after compactification on a small circle. For instance, consider a theory with a single $U(1)$ in $D$ dimensions and just one massive charged particle, with charge $q = 1$. We attempt to construct a strongly self-repulsive multiparticle state of charge $(q,n)=(2,1)$ in $d$ dimensions. However, when $R<R_{\text{crit}}$, the KK modes $(1,0)$ and $(1,1)$ attract each other, and cannot be used together. Likewise, when $R<R_{\text{crit}}'$, the KK modes $(1,2)$ and $(1,-1)$ attract each other and cannot be used together. More generally, any multiparticle state with a charge vector parallel to $(2,1)$ must contain at least one positive KK mode and one non-positive KK mode by charge conservation, but these modes attract each other when both $R<R_{\text{crit}}$ and $R<R_{\text{crit}}'$, and so for small enough $R$ there is no strongly self-repulsive multiparticle state of KK modes in this charge direction, and the RFC can be violated.

More generally, if there are only a finite number of charged particles in $D$ dimensions then we can always find a sufficiently small radius for which the $n_1$ and $n_2$ KK modes of any two massive charged particles in the theory attract each other whenever $n_1 n_2 \le 0$.\footnote{Massless charged particles provide an interesting complication, but we can introduce Wilson lines $\theta^a \ne 0$ to give all charged particles a $d$-dimensional mass, in which case the same conclusion follows whenever $\tilde{n}_1 \tilde{n}_2 < 0$.} In the four dimensional case, for each mutually attractive pair a bound state will form. This bound state may be self-repulsive, thereby satisfying the RFC for this direction in charge space, but it is not guaranteed to be (in examples, such a bound states is often the KK mode of another particle, or able to decay into other KK modes). Thus, the presence of a self-repulsive charged particle in $D$ dimensions is not sufficient to ensure that the RFC will be satisfied after KK reduction.

As in the case of the WGC, the simplest resolution is to demand an infinite tower of self-repulsive particles in $D$ dimensions. We may define the Tower RFC and sub-Lattice RFC accordingly:
\begin{namedconjecture}[The Tower Repulsive Force Conjecture (TRFC)]
Given any site in the charge lattice, $\vec{q} \in \Gamma$, there exists a positive integer $n$ such that there is a self-repulsive particle of charge $n \vec{q}$.
\end{namedconjecture}
\begin{namedconjecture}[The sub-Lattice Repulsive Force Conjecture (sLRFC)]
There exists a positive integer $n$ such that for any site in the charge lattice, $\vec{q} \in \Gamma$, there is a self-repulsive particle of charge $n \vec{q}$.
\end{namedconjecture}
\noindent
Unlike the RFC, both of these conjectures are preserved under (tree-level) KK reduction, whereas sLRFC implies the TRFC, which implies the RFC.
Note the RFC would not follow from either conjecture if we demanded that it be satisfied by stable particles, since the TRFC and sLRFC (like the TWGC and sLWGC) generically require resonances, and even an infinite tower of unstable self-repulsive resonances does not guarantee the existence of a single stable self-repulsive particle. Indeed devising a simple conjecture that implies a stable-particle version of the RFC after KK reduction is surprisingly difficult, another good reason to omit this requirement from the conjecture.

Heterotic string theory compactified to $D \le 10$ dimensions on a torus provides a simple example where both the TRFC and the sLRFC are satisfied (with $n=1$ in either case). Details can be found in appendix~\ref{sec:heterotic}.

\section{WGC vs. RFC}\label{sec:vs}

\subsection{Examples of theories obeying the WGC and the RFC}

As illustrated in Fig.~\ref{fig:phase}, the WGC and the RFC are independent conjectures---either one can, in principle, be satisfied when the other is false. In some contexts, however, they reduce to the same statement. We will give two simple examples of theories in which this happens. In the first case, toroidal compactifications of theories of pure gravity, both bounds are saturated. These theories can be embedded in a supersymmetric setting where the charged particles are BPS states that are both extremal and marginally self-repulsive. We expect that theories where the RFC differs from the WGC will need sufficient supersymmetry to protect the existence of massless scalars, but should have charged particles which are not extremal BPS states. Our second example fits the bill: the 10d heterotic string, for which the particles charged under the gauge group are not BPS. In this case the WGC and the RFC are in principle different. Interestingly, we find that the form of the spectrum implies that they are closely linked to each other. This provides an illustrative example of how the two independent conjectures can be simultaneously satisfied by a simple ansatz for the spectrum.

On the other hand, the WGC and RFC bounds are not always identical. In particular, we will see that for M-theory compactified on a Calabi-Yau three-fold, BPS states that becomes massless at a conifold transition are strictly superextremal but marginally self-repulsive. Thus, these BPS state satisfy both the RFC and the WGC, but the former only marginally, whereas they satisfy the latter with room to spare.

\subsubsection{Toroidal compactifications of pure gravity}

If we compactify $D$-dimensional Einstein gravity on an $r$-torus, we obtain a theory with gauge group U(1)$^r$ and with $r(r+1)/2$ massless moduli fields, parametrizing the size and shape of the torus. (For instance, in the case $r = 2$, we can think of two of the three scalar fields as radions for the two circles, while the third field can be thought of as the axion arising from a Wilson line of the first graviphoton around the second circle.) We can parametrize the moduli in the form of a symmetric matrix of fields $\varphi_{ij}$ with determinant $|\varphi|$.

The necessary formulas for this case are all conveniently summarized in \S2.1 of \cite{Heidenreich:2016aqi}. The Kaluza-Klein modes are labeled by their charges $Q_i$, $i \in \{1, \ldots, r\}$, and have mass
\begin{equation}
m^2(Q) = |\varphi|^{-\frac{1}{d-2}} \varphi^{ij} Q_i Q_j/R^2,
\end{equation}
with $\varphi^{ij}$ the inverse matrix of $\varphi_{ij}$. The metric on scalar field space can be read off from the kinetic term in $d = D-r$ dimensions in Einstein frame,
\begin{equation}
-\frac{1}{2} \int d^d x \,G^{ij,kl} \nabla \varphi_{ij} \cdot \nabla \varphi_{kl} = \frac{1}{2\kappa_d^2} \int d^d x \sqrt{-g} \biggl(-\frac{1}{4} \left[\varphi^{ik}\varphi^{jl} + \frac{1}{d-2}\varphi^{ij}\varphi^{kl}\right] \nabla \varphi_{ij} \cdot \nabla \varphi_{kl}\biggr).
\end{equation}
The inverse metric is then
\begin{equation}
G_{ij,kl} = 4\kappa_d^2 \left[ \varphi_{ik} \varphi_{jl} - \frac{1}{d + r - 2} \varphi_{ij} \varphi_{kl}\right] \quad \Rightarrow \quad G_{ij,mn}G^{mn,kl} = \delta_{ij}^{kl}.
\end{equation}
The scalar force involves the combination $G_{ij,kl} \frac{\partial m(Q)}{\partial \varphi^{ij}} \frac{\partial m(Q)}{\partial \varphi^{kl}}.$ Using standard formulas for the derivative of an element of an inverse matrix or of the determinant of a matrix with respect to entries in the matrix, it is a straightforward exercise to check that each KK mode exactly saturates the RFC inequality. This is to be expected, because if we started with a sufficiently supersymmetric theory in $D$ dimensions, then the Kaluza-Klein modes of the graviton are all BPS particles.

\subsubsection{The heterotic string in 10d}

More interesting examples arise in theories where the charged particles are not BPS states. As an example, consider the heterotic string in 10 dimensions, for which the lightest state of charge $Q_i$ has mass
\begin{equation}
m^2 = \frac{2}{\alpha'} (|Q|^2 - 2) = e^2(\Phi) M_{10}^8 (|Q|^2 - 2).
\end{equation}
The modulus is the dilaton $\Phi$, with string coupling $g_s = \exp(\Phi)$, and we have used the two relations $e^2 = g_s^2 (2\pi)^7 \alpha'^3$ and $M_{10}^{-8} = \frac{1}{2} g_s^2 (2\pi)^7 \alpha'^4$. The familiar WGC bound in this case is given by 
\begin{equation}
e^2 |Q|^2 M_{10}^8 \geq \left[\frac{\alpha^2}{2} + \frac{7}{8}\right] m^2,
\label{eq:10dheteroticWGC}
\end{equation}
where the gauge kinetic term contains a prefactor $\e^{-\alpha \Phi}$ and in the heterotic string $\alpha = \frac{1}{2}$. (See \cite{Heidenreich:2015nta} for a more complete discussion of the heterotic string in our conventions.)

The RFC bound takes a similar form, multiplying through by $M_{10}^8$:
\begin{equation}
e^2 |Q|^2 M_{10}^8 \geq 2 \left(\frac{\partial m}{\partial \Phi}\right)^2 + \frac{7}{8} m^2.
\label{eq:10dheteroticRFC}
\end{equation}
Recall that the derivative is taken at fixed $M_{10}$. We can use the relation $\alpha'^4 = \frac{2}{\exp(2\Phi) (2\pi)^7 M_{10}^8}$ to compute $d\alpha'/d\Phi = -\alpha'/2$, leading to:
\begin{equation}
\frac{dm}{d\Phi} = \frac{1}{2m} \frac{dm^2}{d\alpha'} \frac{d\alpha'}{d\Phi} = \frac{1}{2m} \left(-\frac{1}{\alpha'} m^2\right)\left(-\frac{\alpha'}{2}\right) = \frac{m}{4}.
\end{equation}
But then, the $2 \left(\frac{\partial m}{\partial \Phi}\right)^2$ term in \eqref{eq:10dheteroticRFC} reduces to $m^2/8$, which exactly matches the $\alpha^2 m^2/2$ term in \eqref{eq:10dheteroticWGC}.

This calculation shows that, despite not being BPS states, the charged particles in the 10d heterotic string spectrum obey the RFC. This is true for more general theories with dilatonic couplings \cite{Lee:2018spm}, and it is also true for toroidal compactifications of the heterotic string, as we show in appendix \ref{sec:heterotic}. In fact, given the dependence of the particle masses on the moduli fields, the RFC reduces to {\em precisely} the same inequality that the WGC does. This somewhat surprising result is a consequence of the factorized form of the spectrum: as noted in \cite{Lee:2018spm}, if for a conventionally normalized modulus field $\phi$ coupled to a gauge field kinetic term with a factor $\exp(-\alpha \phi)$ we have a spectrum
\begin{equation}
m^2(Q) = e^2(\phi) M_{d}^{d-2} f(Q) = e^2(0) \exp(\alpha \phi) M_{d}^{d-2} f(Q),
\end{equation}
then the RFC will always take the form
\begin{align}
e^2 |Q|^2 M_{d}^{d-2} \geq 2 \left(\frac{\partial m}{\partial \phi}\right)^2 + \frac{d-3}{d-2} m^2 
= \left[\frac{1}{2} \alpha^2 + \frac{d-3}{d-2}\right] m^2,
\end{align}
which is the WGC bound.\footnote{More generally, one can show that in any two-derivative theory of moduli, gauge fields, and gravity, a particle that is self-repulsive everywhere in moduli space is superextremal, and a particle that has zero self-force and nonzero mass everywhere in moduli space is extremal~\cite{BHpaper}.}

Such simple spectra are clearly not universal, but it is plausible that spectra at asymptotically weak coupling will often take this form, as suggested by the Swampland Distance Conjecture \cite{Ooguri:2006in}.

\subsubsection{M-theory on the conifold}

In some cases, BPS bounds and extremality bounds coincide. This happens in the first example we considered: Kaluza-Klein modes of pure gravity on a torus are both BPS and extremal in theories with sufficient supersymmetry. In some cases, they do not agree. This happens for the second example we considered: in heterotic string theory on a torus, extremal black holes with $Q_L^2 > Q_R^2$ will not be BPS.\footnote{For instance, black holes that are predominantly charged under the $E_8 \times E_8$ or $\mathrm{Spin}(32)/\mathbb{Z}_2$ gauge group---or the $U(1)^{16}$ that remains after turning on generic Wilson lines---will satisfy this condition.} 

In 5d $\mathcal{N}=1$ supergravity theories, there is a simple criterion for determining when the BPS bound will coincide with the extremality bound in a given direction in charge space, so that BPS $\equiv$ extremal: this happens if and only if the central charge of a state in this charge direction does not vanish anywhere in moduli space \cite{conifolds} (see also~\cite{BHpaper}).\footnote{In 4d $\mathcal{N}=2$ theories, the relationship between the BPS bound and the extremality bound is likely related to the ``type" of the BPS state discussed in \cite{Grimm:2018cpv,Grimm:2018ohb}.}

A common instance in which the central charge does vanish is the conifold. In M-theory on a resolved conifold geometry, there is a charged BPS state associated with an M2-brane wrapping an $S^2$ that can shrink to zero size, forming a conifold singularity. When this happens, the central charge of this BPS state vanishes, and the state (a hypermultiplet) becomes massless. As a result, the BPS bound and the extremality bound for this BPS state do not coincide: the state is BPS but strictly superextremal. Since BPS states always have vanishing self-force, this state saturates the RFC bound but satisfies the WGC bound with room to spare.

A similar phenomenon occurs in heterotic string theory compactified on a torus. The spectrum is determined by an even self-dual lattice $\Gamma$, and the lightest charged particle for a given $Q = (Q_L, Q_R) \in \Gamma$ has various properties, depending on $Q^2 \df Q_L^2 - Q_R^2 \in 2\mathbb{Z}$. Either (1) $Q^2 \le 0$, and the lightest charged particle is both BPS and extremal, or (2) $Q^2 = 2$, and the lightest charged particle is BPS, but strictly superextremal, or (3) $Q^2 > 2$, and the lightest charged particle is strictly superextremal, yet non-BPS. This is similar to the M-theory examples just discussed, but the BPS states that become massless at special points in the moduli space (those with $Q^2=2$) are vector multiplets ($W$ bosons), and the gauge symmetry has a non-Abelian enhancement when this occurs.

From our analysis thus far, it is clear why the BPS bound and extremality bound do not necessarily coincide, even though extremal black holes have vanishing self-force: a particle can satisfy a zero self-force condition yet be strictly superextremal if the scalar force acts more strongly on it than it does on an extremal black hole. Similarly, a particle could satisfy a zero self-force condition yet be strictly subextremal if the scalar force acts more weakly on it than it does on an extremal black hole.\footnote{However, the latter cannot be true everywhere in moduli space~\cite{BHpaper}, which is perhaps unsurprising, since BPS states cannot be subextremal.}

In the case of BPS states, the BPS bound ensures that no particle can have a larger charge-to-mass ratio than a BPS state in the charge direction of interest. This means that BPS states can only be extremal or superextremal, as extremal black holes would violate the BPS bound if BPS states were subextremal. In other words, BPS states feel a scalar self-force that is at least as strong (relative to their charge) as a black hole in their direction in charge space. The BPS bound and extremality bound agree if and only if the scalar charge-to-mass ratio $\mu/m$ is the same for a BPS state as it is for an extremal black hole.

\subsection{Why the WGC and RFC are related}\label{ssec:related}

In theories without massless scalar fields mediating a long-range force, self-repulsiveness and superextremality become equivalent, as do the WGC and RFC. Once scalar fields are allowed in the game, however, the situation becomes more complicated.

Figure \ref{fig:phase} shows that in such a theory, a particle can be superextremal yet self-attractive, or self-repulsive yet subextremal. One might suspect, therefore, that a theory could in principle satisfy the RFC but not the WGC, or vice versa. Indeed, it is not hard to imagine a scenario in which the RFC is satisfied but the WGC is not: simply take a spectrum that violates the WGC and add a single self-repulsive subextremal particle. This is possible when the particle couples to a massless scalar more weakly than extremal black holes do.

More interesting consequences follow if the massless scalars are moduli and the RFC is satisfied everywhere in moduli space. In particular, if we make the seemingly minimal assumption that a particular particle species is self-repulsive everywhere in the moduli space, then it turns out that this particle must also be \emph{superextremal} everywhere in moduli space~\cite{BHpaper}, and the WGC follows! Likewise, if any fixed multiparticle state of charge $\vec{Q}$ is weakly self-repulsive everywhere in moduli space, then it is superextremal everywhere in moduli space, and the WGC is satisfied in the direction of $\vec{Q}$.

Thus, to violate the WGC and satisfy the RFC, there must be multiple particles and/or multiparticle states that are self-repulsive in distinct regions of moduli space. Depending on the coupling of the modulus to gauge fields, the number of distinct regions with different self-repulsive particle content required to satisfy the RFC everywhere in moduli space without satisfying the WGC \emph{almost} everywhere in moduli space could be infinite; for instance, this is true for a dilaton. Therefore, while no clear inconsistencies result, satisfying the RFC across moduli space without also satisfying the WGC places interesting, nontrivial constraints on the theory.

The converse possibility of a theory that satisfies the WGC but not the RFC is more bizarre, at least in 4d. Pick any rational charge direction in which the RFC is violated, and consider a superextremal multiparticle state in this direction of mass $m_1$, charge $\vec{q}_1$ and charge-to-mass ratio $\vec{z}_1 = \vec{q}_1/m_1$. The argument proceeds as in~\S\ref{subsec:RFC}: by assumption, given two copies of this multiparticle state, some pair of particles will be mutually attractive---otherwise the multiparticle state would be strongly self-repulsive, and the RFC would be satisfied in this direction. Allowing these particles to bind together and the bound state to decay if unstable, we obtain a new multiparticle state with charge $\vec{q}_2 = 2 \vec{q}_1$ and mass $m_2 < 2 m_1$ (due to the binding energy and any kinetic energy released by the decay). Thus, the charge-to-mass ratio has increased, $|\vec{z}_2| > |\vec{z}_1|$.

Iterating, we obtain multiparticle states with every increasing charge-to-mass ratios $|\vec{z}_3| > |\vec{z}_2|$, $|\vec{z}_4| > |\vec{z}_3|$, etc. This is the usual consequence of violating the RFC in 4d, but now the multiparticle states are all \emph{superextremal}. In particular, assuming a finite number of stable particles below any given mass scale, this implies an infinite tower of charged particles with ever-increasing, superextremal charge-to-mass ratios.

This is not quite the TWGC we have defined above, as the charge sites populated by these superextremal particles could be very sparse. Selecting $\vec{q} = 3 \vec{q}_1$, for instance, the above argument does not ensure the existence of some integer $n$ with a superextremal particle of charge $n \vec{q}$. Nonetheless, we are guaranteed an infinite tower of superextremal states in each direction in the charge lattice for which no self-repulsive state exists, which has the same flavor as the TWGC.

Furthermore, the requirement that this infinite tower of states must have an \emph{increasing} charge-to-mass ratio is quite 
unusual from the perspective of black hole physics. Essentially by definition, at large charge, we should have a black hole spectrum with charge-to-mass ratios that asymptote to $|\vec{Z}_{\rm BH}|$, either from above or below. In the former case, the charge-to-mass ratios \emph{decrease} as the charge of the black holes goes to infinity. In the latter case, all of these finite-sized black holes are slightly \emph{subextremal}. In neither case do we see a tower of superextremal black hole states with increasing charge-to-mass ratio.

This means that the tower of states implied by this reasoning must not be black hole states. This does not necessarily present a problem: one could imagine that our tower of superextremal states involves weakly-bound objects, with a radius that is much larger than their Schwarzschild radius. While there is no sharp inconsistency with this outcome that we are aware of, it seems pathological. To avoid it, one must insist that the tower of states terminates on a self-repulsive state, so the RFC is satisfied. We conclude that, aside from the strange situation described above, a 4d theory that satisfies the WGC must also satisfy the RFC.

\subsection{On unifying the two conjectures} \label{sec:MZC}

We have seen that the WGC and RFC are distinct conjectures, with neither one necessarily implying the other. However, they are closely related, and become equivalent in the absence of massless scalars. Even with massless scalars, the WGC and RFC are very similar, and violating one while preserving the other has some unexpected consequences, as described above. To the best of our knowledge, there are no known counterexamples to either.

It is somewhat surprising that two such closely related conjectures should remain distinct, with both (apparently) satisfied in all known examples of quantum gravities.
It is interesting, therefore, to consider whether they can be elegantly unified into a single conjecture, implying both of them. In four dimensions, the following conjecture, which we call the Maximal $Z$ Conjecture (MZC), does the job:
\begin{namedconjecture}[The Maximal $Z$ Conjecture (MZC)]
For every rational direction in charge space, there exists a multiparticle state of maximal $| \vec{Z} | \df |\vec{Q}|/m$.
\end{namedconjecture}
\noindent
Why does this imply the WGC? In theories that violate the WGC, extremal black holes of finite size will be slightly subextremal, and kinematically there will be a infinite tower of stable black hole states of increasing $Z$ in some direction in charge space, which asymptotes to the charge-to-mass ratio $Z_{\text{ext}}$ of an infinitely-large black hole. Since the WGC is violated by assumption, every state in the theory has $Z < Z_{\text{ext}}$, and therefore no state has $Z$ larger than or equal to all other states, and the MZC is violated.

The converse is not true: the WGC does not imply the MZC, as one could imagine an infinite tower of weakly-bound superextremal states of increasing $Z$, as previously considered in \S\ref{ssec:related}. Note, however, that if the convex hull is generated by a finite number of stable particles, then both the WGC and the MZC are satisfied. None of these arguments are specific to $D=4$; the MZC is strictly stronger than the WGC in a general number of spacetime dimensions.

The MZC also implies the RFC, but only in 4d: suppose that the RFC is violated in a particular charge direction. By assumption, any multiparticle state in this direction is self-attractive. Taking two copies of the state, we form a new multiparticle state with larger $Z$ by allowing a mutually attractive pair of particles to bind together. Thus, no multiparticle state in this direction can have maximal $Z$, and therefore a 4d theory that violates the RFC also violates the MZC. The same argument does not work in $D\ge 5$ dimensions because mutually attractive particles do not always form bound states.

In defining the RFC, we argued that it is most natural to allow unstable, narrow resonances to satisfy the conjecture. By comparison, for the MZC (like the WGC) this is a moot point: a particle of maximal $Z$ is either kinematically stable or can only decay at threshold to a multiparticle state with the same $Z$. This is because charge and energy conservation do not allow $Z$ to decrease in a decay process; if kinetic energy is released, then $Z$ must increase, whereas decays at threshold (such as wall-crossing phenomena) leave $Z$ unaltered.

Similarly to the WGC and RFC, a theory that satisfies the MZC before compactification can violate it afterwards. However, unlike the WGC and RFC, defining suitable strong forms of the MZC is a difficult task. One would like to define the Tower MZC---by analogy with the TWGC and TRFC---as a statement about unstable particles of maximal charge-to-mass ratio amongst all particles far out in the charge lattice. But in this case, there are subtleties in relating statements about single and multiparticle states. Moreover, it is unclear whether such a Tower MZC would be preserved under dimensional reduction. Thus, we refrain from positing any particular strong form of the MZC, but note that a stronger condition than the MZC itself must be satisfied in $D$ dimensions to ensure that the MZC is satisfied after dimensional reduction.

\section{The RFC in non-gravitational theories} \label{sec:RFCnongrav}

The WGC bound, $q/m \geq 1/M_{D}^{(D-2)/2}$, is manifestly a statement about gauge theories coupled to gravity. When gravity is decoupled, $M_D \rightarrow \infty$, the bound becomes trivial.

The RFC, on the other hand, is a meaningful statement even after gravity is decoupled. Assuming that the self-repulsive particles do not also decouple in this limit, the RFC would imply that in a UV-complete quantum field theory, for every direction $\hat{q}$ in the charge lattice, there must exist a self-repulsive charged (possibly multiparticle) state. As with the quantum gravity swampland, this would only apply to UV-complete theories (such as asymptotically free theories); to distinguish UV-complete theories from general effective field theories is the main goal of the swampland program.

However, nothing obviously prevents the self-repulsive particles from decoupling, and indeed it is trivial to violate the above, naive conjecture: a free Abelian gauge theory is UV-complete, but contains no charged particles and therefore no self-repulsive particles.
Intriguingly, a minimal modification of the RFC designed to exclude this trivial counterexample is much harder to disprove:
\begin{namedconjecture}[The Repulsive Force Conjecture for Quantum Field Theories (RFC-for\-QFTs)]
For every direction in charge space in which there is a charged multiparticle state, there is a strongly self-repulsive multiparticle state.
\end{namedconjecture}
\noindent
We have not yet found a definitive counterexample to this statement. Below we discuss a potential 4d counterexample in which the \emph{elementary} charged particles do not quite satisfy the conjecture in the form stated above. However, by the same reasoning as in the gravitational case, bound states will form, and it is more difficult to determine whether they will fill in the gaps.

A counterexample necessarily requires massless scalars to mediate a self-attractive force stronger than the gauge force. Thus, any counterexamples (if they exist) are likely to be supersymmetric theories with moduli.

It is worth noting that the RFC-for-QFTs is preserved under dimensional reduction. In the gravitational case, we saw that a violation of the WGC and RFC can occur upon circle reduction unless the parent $D$-dimensional theory has an infinite tower of charged states. In the non-gravitational case, however, UV-complete quantum field theories do not require an infinite number of particles. Without such a tower, na\"ively one might worry that KK reduction would lead to a violation of the RFC-for-QFTs in $d$ dimensions. However, because gravity is non-dynamical, the radion and Kaluza-Klein photon are also non-dynamical, and do not mediate long-range forces. Thus, a self-repulsive state in the parent theory necessarily descends to a self-repulsive state in the daughter theory, and there is no risk of violating the CHC after such a reduction.

It is not clear why this conjecture should be true. In four dimensions, a heuristic argument in its favor is as follows: suppose the RFC-for-QFTs is violated, and there is a self-attractive particle of charge $\vec{q}$. We can then form a bound state between such particles, with total charge $2 \vec{q}$, which by assumption must also be self-attractive. Iterating, we get a whole tower of particles with increasing charge-to-mass ratio. One might worry that, allowing these particles to run in loops, we generate a positive $\beta$-function for the gauge coupling, thereby precluding a UV-completion, even if the $U(1)$ completes to a non-Abelian gauge group above some scale. However, this argument is not convincing because these states may be very weakly bound, thereby contributing insignificantly to the $\beta$-function. (In~\S\ref{sec:vs}, this same loophole prevented us from concluding that the WGC implies the MZC or the RFC in gravitational theories.)

The MZC is also well-defined for non-gravitational theories, with the same caveats about free Abelian gauge fields as before. By the same argument as in the gravitational case, a four-dimensional QFT that satisfies the MZC-for-QFTs must also satisfy the RFC-for-QFTs.

One reason for taking interest in these non-gravitational conjectures is that they measure the extent to which the Weak Gravity Conjecture should be viewed as an intrinsically gravitational phenomenon. It could be, for instance, that the MZC is a universal property of both UV-complete quantum field theories and quantum gravities. If so, the WGC would follow as an immediate consequence.

\subsection{Potential counterexamples on 4d \texorpdfstring{$\mathcal{N} = 2$}{N=2} Coulomb branches}

The Coulomb branch of a 4d $\mathcal{N}=2$ gauge theory is a simple, well-controlled setting in which we can test the RFC-for-QFTs. In particular, we focus on Coulomb branches of pure glue theories. For some choices of gauge group, such as $\SU{2}$ or $\SU{3}$, the conjecture (restricted to electric charges) is satisfied by the $W$ bosons (which are BPS states), at least when we are far out on the Coulomb branch. However, for other choices---the simplest being $\SU{4}$---the $W$ bosons themselves do not quite satisfy the conjecture. Although they are self-repulsive (being BPS), they do not form strongly self-repulsive multiparticle states in every charge direction.

At a generic point on the Coulomb branch, the central charges of any two distinct $W$ bosons do not align, and the force between them is nonvanishing. In general, the mutual force between a BPS state of charge $\vec q$ and another of charge $\vec{q}\,'$ is given by
\begin{equation}
\cF_{12} = 4 \pi t^{ij} q_i q_j' \text{Re} \left[ 1 - \frac{\bar Z_{\vec{q}} Z_{\vec{q}\,'} }{ | Z_{\vec{q}} Z_{\vec{q}\,'}| }    \right],
\label{eq:pureglueselfforce}
\end{equation}
where $t_{ij} = \text{Im } \tau_{ij}$ is the imaginary part of the gauge kinetic matrix, $t^{i j}$ is its inverse, and $Z_{\vec{q}} = \vec{q} \cdot \vec{a}$ is the central charge of a state of charge $\vec{q}$, where $\vec{a}$ are the Coulomb branch parameters. The mass of the BPS state is simply $m_{\vec{q}} = |Z_{\vec{q}}|$.
 Far out on the Coulomb branch, the gauge kinetic matrix is determined by asymptotic freedom and the one-loop beta function
\begin{equation}
\tau_{ij}(\vec a) = \frac{2i}{\pi} \sum_{\alpha > 0 } q_i^\alpha q_j^\alpha \log \left( \frac{\vec{a} \cdot \vec{q}\,^\alpha}{\Lambda} \right) + ...,
\label{eq:tauij}
\end{equation}
where $\Lambda$ is a dynamically-generated scale and the sum runs over the positive roots of the gauge group, each of charge $\vec{q}\,^\alpha$.

From (\ref{eq:pureglueselfforce}), we see that the self-force of a $W$ boson of charge $\vec{q}$ will vanish, as the term in brackets vanishes. However, if the phases the central charges of two distinct $W$ bosons differ, then the term in brackets will be strictly positive, and the question of whether or not the force is repulsive depends on whether or not the inner product of the charges, computed with respect to $t^{ij}$, is positive or negative. This question can be answered by computing $\tau_{ij}$ via (\ref{eq:tauij}), and in general the answer is moduli-dependent.

\begin{figure}
\centering
\includegraphics[width=6cm]{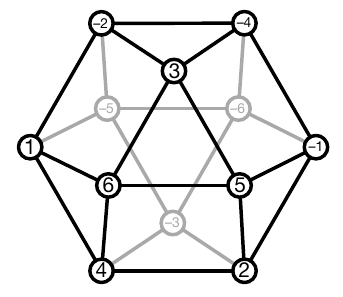}\
\caption{The root system of the group SU(4) with the positive roots labeled 1--6. The negative root associated to a positive root $n$ is denoted $-n$. Roots connected by solid black lines are $60^\circ$ apart, whereas those diagonally opposite on the square faces are $90^\circ$ apart, and all other angles are obtuse.}
\label{fig:SU4roots}
\end{figure}

The root system of $SU(4)$ is shown in figure \ref{fig:SU4roots}, with the six positive roots labeled 1--6. One can show that the inner product between pairs of roots connected by a solid black line is necessarily non-negative, so these $W$ bosons are mutually repulsive everywhere far out on the Coulomb branch ($a_i \gg \Lambda$). On the other hand, roots that are diagonally opposite on a square face of the polytope do not have a definite sign inner product: the $W$ bosons $W_1$ and $W_3$ associated with roots $\alpha_1$ and $\alpha_3$ are mutually repulsive in some regions of moduli space and attractive in other regions, whereas $W_1$ and $W_{-3}$ are mutually attractive when $W_1$ and $W_3$ are mutually repulsive, and vice versa. The same is true for the $W$ bosons pairs $W_4$ and $W_5$ as well as $W_2$ and $W_6$.

Using (\ref{eq:tauij}), one can show that there are regions in moduli space for which both pairs $W_2$ and $W_6$ as well as $W_4$ and $W_5$ are mutually attractive. In these regions there are no strongly self-repulsive multiparticle states directed into the interior of the square face surrounded by these roots (for instance, in the charge direction $q^{\alpha_2}+ q^{\alpha_6}=q^{\alpha_4}+ q^{\alpha_5}$). The same is true for the other square faces in figure \ref{fig:SU4roots}, but in disjoint regions of the moduli space.
 Thus, although their strongly self-repulsive multiparticle states cover most directions in charge space, the $W$ bosons alone are not sufficient to satisfy the RFC-for-QFTs. However, this does not exclude the possibility that some bound state of these $W$ bosons (or a bound state of bound states) could be strongly self-repulsive, so it not obvious whether this theory is a counterexample to the RFC-for-QFTs. 

\section{Conclusions}\label{sec:CONC}

In this work, we have compared and contrasted two distinct but closely related conjectures, both of which have previously been considered in the literature. After reviewing some of the arguments in favor of the WGC, we saw that many of these arguments (consistency under dimensional reduction, examples in string theory, gauge-scalar-gravity unification) can also be used to argue in favor of the RFC as well. We also saw that the consistency issues that plague the WGC under Kaluza-Klein compactification also arise for the RFC, which motivated us to consider strong forms of the conjecture. In four dimensions, we noted that the WGC and RFC both follow from a conjecture that we called the ``Maximal $Z$ Conjecture (MZC),'' which holds that in any direction in the charge lattice, there is a state of maximal charge-to-mass ratio.

Each of these conjectures has its own advantages. The WGC depends on the notion of ``superextremality,'' which has the advantage of being preserved under the formation of bound states and the decay of unstable particles, so the lightest superextremal state is stable. Likewise, the lightest state of maximal charge-to-mass ratio is stable, so if either the MZC or WGC is true, they are necessarily satisfied by stable states. On the other hand, the RFC depends on the notion of ``self-repulsiveness,'' which is not necessary preserved under bindings or decays. As a result, it is natural to include unstable, narrow resonances in the conjecture. While a strictly stronger conjecture demanding stable self-repulsive states can be formulated, this conjecture is more difficult to verify or refute in examples, and we have not considered it here.

Another advantage of the WGC is that it connects more closely to black hole physics, one of the principal motivations for the conjecture in the first place. An extremal black hole can decay only by emitting superextremal particles. In a theory without massless scalars, these particles are necessarily self-repulsive, but in the presence of massless scalars, they need not be. Thus, the decay of black holes is intimately connected to the WGC, whereas its relationship with the RFC and MZC is less obvious.

On the other hand, one advantage of the RFC and the MZC is that they are not inherently gravitational, so they can be formulated in UV-complete quantum field theories (without gravity). We have seen an example of an $\mathcal{N}=2$ theory in which the elementary particles do not satisfy the RFC-for-QFTs, but it is possible that bound states may satisfy the conjecture. We defer a more thorough investigation to future work.

A number of potential research directions have arisen in the course of this work. The effects of higher-derivative operators on the black hole extremality bound in the absence of massless scalars have been studied in some detail. However, the effect of massless scalars on both extremality and self-repulsiveness beyond the leading two-derivative action is presently unknown. Determining this would be very useful for future studies of both the WGC and RFC, particularly in supersymmetric contexts.

While we presented a number of arguments and examples in support of the RFC, we did not attempt to extend every argument for the WGC to an analogous statement about the RFC. It would be interesting to see if there is any sort of relationship between the RFC and black hole entropy, infrared consistency, modular invariance, or cosmic censorship. While the RFC and WGC are essentially identical in the absence of massless scalars, the most precise tests of either are in supersymmetric examples, where moduli are ubiquitous. Thus, there is a good chance that the nature of their relationship will become clearer after further investigations into both.

\section*{Acknowledgements}

We thank Nima Arkani-Hamed, Clay C\'ordova, Finn Larsen, Mehrdad Mirbabayi, Kantaro Ohmori, Eran Palti, and Irene Valenzuela for helpful discussions. The research of B.H.~was supported in part by NSF grant PHY-1914934 and in part by Perimeter Institute for Theoretical Physics. Research at Perimeter Institute is supported by the Government of Canada through the Department of Innovation, Science and Economic Development, and by the Province of Ontario through the Ministry of Research, Innovation and Science. M.R.~is supported in part by the DOE Grant DE-SC0013607. 
T.R.~is supported by the Carl P. Feinberg Founders Circle Membership and by NSF grant PHY-1606531. The authors thank the Simons Center for Geometry and Physics and the Simons Summer Workshop for hospitality in the early stages of this research. B.H.~thanks the Yau Mathematical Sciences Center and the Kavli Institute for the Physics and Mathematics of the Universe for hospitality during the final stages of this work. This work was completed at Aspen Center for Physics, which is supported by NSF grant PHY-1607611.

\appendix

\section{Bound states in diverse dimensions}\label{sec:bound}

Classically, any pair of mutually attractive particles in $D=d+1$ spacetime dimensions will attract each other and form a bound state. Quantum-mechanically, however, zero point energy can disrupt the bound state, and consequently its existence is not guaranteed (see, e.g.,~\cite{hydrogenatoms}).

Finding non-relativistic bound states amounts to finding normalizable solutions to a hydrogen atom-like time-independent Schr\"odinger equation in $d$ spatial dimensions:
\begin{align}
\left[ -\frac{\hbar^2}{2\mu} \nabla_d^2 - \frac{\alpha}{r^{d-2}} \right] \Psi(\bm{r}) = E \Psi(\bm{r}),
\end{align}
where $\nabla_{d}^2 = \frac{\partial^2}{\partial r^2} + \frac{d-1}{r} \frac{\partial}{\partial r} + \frac{1}{r^2} \Omega^2$ is the $d$-dimensional Laplacian, $\Omega^2$ is the $(d-1)$-sphere Laplacian, $\mu = m_1 m_2/(m_1+m_2)$ is the reduced mass and $\alpha$ is the ``fine structure constant,'' $\alpha = -\frac{\mathcal{F}_{12}}{(d-2)V_{d-1}}$. We emphasize that, in this appendix, $d$ is the number of \emph{spatial} dimensions (not the number of spacetime dimensions after compactification) and $\mu$ is the reduced mass (not the scalar charge). This differs from our conventions in the main text.

We assume $\alpha > 0$ (mutual attraction), so that classical bound states can form. Expanding the wavefunction in $d$-dimensional spherical harmonics, we obtain the radial equation:
\begin{equation}
\frac{d^2 R(r)}{dr^2} + \frac{d-1}{r}\frac{dR}{dr}+\frac{2\mu}{\hbar^2}\left[E + \frac{\alpha}{r^{d-2}} - \frac{\hbar^2}{2 \mu} \frac{ \ell (\ell+d-2) }{r^2} \right] R = 0 \,,
\end{equation}
for each harmonic, where the spherical harmonics of order $\ell$ are restrictions of homogeneous degree-$\ell$ harmonic polynomials on $\mathbb{R}^d$ to the surface of the sphere.
Setting $u(r) = r^{(d-1)/2}R(r)$, we obtain
\begin{equation}
\frac{d^2 u(r)}{d r^2} + \frac{2 \mu}{\hbar^2}\left[ E-V_{\rm{eff}}(r)\right] u(r) = 0,
\end{equation}
with
\begin{equation}
V_{\rm{eff}}(r) = -\frac{\alpha}{r^{d-2}} +\frac{\hbar^2}{2 \mu} \frac{j(j+1)}{r^2}\,,~~~~j=\ell+\frac{d-3}{2}.
\end{equation}
In the familiar case of $d=3$, the second term in the effective potential---the ``centrifugal barrier''---is proportional to $\ell$, so it vanishes in the absence of angular momentum. However, for $d > 3$ this term is nonzero even for $\ell = 0$, and gives a positive contribution to the effective potential. As a result, whereas bound states with negative energy are guaranteed to exist in $d=3$, they might not exist in higher dimensions due to the centrifugal term \cite{hydrogenatoms}. Note that this is manifestly a quantum effect, which vanishes in the $\hbar \rightarrow 0$ limit.

The question of whether or not the barrier term will prevent the existence of bound states in higher dimensions depends on $\alpha$, $\mu$, and the dimensionality of spacetime. In particular, when $d=4$, the barrier term and the potential $V(r)$ both have a $1/r^2$ dependence. Thus, if
\begin{equation}
\alpha > \frac{3 \hbar^2}{8 \mu}, \label{eqn:d4boundcondition}
\end{equation}
then the long range force overpowers the zero-point energy contribution, and a bound state will form. For smaller $\alpha$, the zero-point energy wins, and the attractive long-range forces do not create a bound state.

To be precise, the $1/r^2$ potential leads to a continuous spectrum of energy eigenstates with arbitrarily negative energies (the bound state problem is scale-invariant). The unbounded-from-below spectrum is due to the small $r$ behavior of the potential, but this is irrelevant in the present context, because short-range forces will contribute to the potential in the $r \to 0$ limit. We cannot compute the energy of the resulting ground state without understanding the short-range forces, but the inequality~\eqref{eqn:d4boundcondition} is sufficient to ensure that some negative energy bound state exists.

In $d > 4$, the barrier term is dominant at large $r$, whereas the Coulomb potential is dominant at small $r$. This is the opposite of the behavior we are used to in $d=3$: the barrier term confines bound states, making them smaller rather than larger. In fact, for $d>4$, the Coulomb potential $-\alpha/r^{d-2}$ allows arbitrarily small bound states (with arbitrarily negative energy) because for a small bound state of size $L$, the zero-point energy $\sim \hbar^2/L^2$ is subdominant to the potential energy $\sim - \alpha/L^{d-2}$. As above, short-range forces will enter at some point and make the maximum binding energy finite and the spectrum discrete.

However, because the barrier term places an \emph{upper bound} on the size of the bound state, it may happen that short range forces become important before a bound state can form. 
We can estimate the maximum size of a negative energy bound state as the radius $r_0$ at which the effective potential $V_{\rm eff}$ passes through zero, giving:
\begin{equation}
r_0^{d-4} = \frac{8 \mu \alpha}{\hbar^2 (d-1)(d-3)} \,.
\end{equation}
For macroscopically-large objects, the right-hand side is large, and there is no problem. However, for particles with sub-Planckian masses $m_1, m_2 < M_D$ and gravitational-strength interactions, $\mathcal{F}_{12} \sim G_N m_1 m_2$, the right-hand side is sub-Planckian, and the computation is untrustworthy. 

In summary, a pair of mutually attractive particles is not guaranteed to form a bound state for $D = d+1 > 4$. 

\section{Toroidal compactification of the heterotic string}\label{sec:heterotic}

In this appendix, we show that the RFC is satisfied in toroidal compactifications of the heterotic string. In fact, the RFC bound here is exactly equivalent to the WGC bound, so the two conjectures become equivalent to each other. This result relies on a remarkable factorization of the self-force into left- and right-moving terms. This complements the work of \cite{Lee:2018spm}, which found that the RFC is satisfied for heterotic compactifications to 6 dimensions on K3 manifolds.\footnote{Since toroidal compactifications introduce additional massless moduli not present in K3 compactifications, arguments for the RFC in the context of heterotic K3 compactifications do not immediately imply the RFC for heterotic toroidal compactifications, nor do our arguments here imply the RFC for heterotic K3 compactifications.}

Consider heterotic string theory compactified down to $D$ dimensions on a $10
- D$ torus. The mass spectrum is
\begin{equation}
  \frac{\alpha'}{4} m^2 = \frac{1}{2} Q_L^2 + N - 1 = \frac{1}{2} Q_R^2 +
  \tilde{N},
\end{equation}
where $N, \tilde{N}$ are non-negative integers and $(Q_L, Q_R) \in \Gamma$ for $\Gamma$ an even-self-dual lattice of
signature $(26 - D, 10 - D)$. Any such lattice can be written as a boost of
some fixed reference lattice $\Gamma_0$:
\begin{equation}
  \Gamma = \Lambda \Gamma_0,
\end{equation}
where $\Lambda \in \mathrm{SO} (26 - D, 10 - D)$ encodes the moduli in the form
of the coset:
\begin{equation}
  \frac{\mathrm{SO} (26 - D, 10 - D)}{\mathrm{SO} (26 - D) \times \mathrm{SO} (10 -
  D)},
\end{equation}
up to discrete identifications.

The extremality bound was already worked out by Sen \cite{Sen:1994eb}:
\begin{equation}
  \frac{\alpha'}{4} m^2 \geq \frac{1}{2} \max (Q_L^2, Q_R^2) .
\end{equation}
We now work out the mutual force, $\mathcal{F}_{12}$. The moduli are those
encoded by $\Lambda$ as well as the $D$-dimensional ``string coupling,''
\begin{equation}
  g_D \df \sqrt{\frac{2 \kappa_D^2}{\ell_s  (2 \pi \ell_s)^{D - 3}}}\,,
\end{equation}
where $\ell_s \df \sqrt{\alpha'}$ is the string length. The usual, ten-dimensional string coupling $g_s$ is not $T$-duality
invariant, so it is more natural to consider $g_D$, where compactification on
a torus of volume $(2 \pi R)^{10 - D}$ gives $g_D \df g_s (\ell_s /R )^{\frac{10 - D}{2}}$.
For example, $g_9 = \sqrt{g_s g_s'}$, where $g_s$ and $g_s' = g_s \ell_s / R$
are the T-dual string couplings.

Holding $\Gamma$ fixed keeps $R$ fixed in string units, so in this case
varying $g_D$ and $g_s$ are equivalent. The $D$-dimensional Einstein-Hilbert
plus dilaton action is
\begin{equation}
  S = \frac{1}{2 \kappa_D^2} \int d^D x\, e^{- 2 \Phi_D}  [\mathcal{R}_S + 4
  (\nabla \Phi_D)^2] = \frac{1}{2 \kappa_D^2} \int d^D x \left[ \mathcal{R}-
  \frac{4}{D - 2}  (\nabla \Phi_D)^2 \right] .
\end{equation}
Thus, $G_{\Phi_D \Phi_D} = \frac{4}{\kappa_D^2  (D - 2)}$. We have
\begin{gather}
  0 = \frac{\partial}{\partial \Phi_D}  \left( \frac{\alpha'}{4} m^2 \right) =
  \frac{\alpha'}{2} m \frac{\partial m}{\partial \Phi_D} + \frac{1}{4} m^2 
  \frac{\partial \alpha'}{\partial \Phi_D} = \frac{\alpha'}{2} m
  \frac{\partial m}{\partial \Phi_D} - \frac{1}{D - 2} \alpha' m^2 \nonumber \\
  \Longrightarrow \qquad  \frac{\partial m}{\partial \Phi_D} = \frac{2}{D
  - 2} m,
\end{gather}
where we use $\alpha' = (2\pi)^{-\frac{2(D-3)}{D-2}} (g_D^2/2)^{-\frac{2}{D-2}}  M_D^{-2} \propto g_D^{- \frac{4}{D - 2}}$ in $D$-dimensional
Planck units. Thus, combining the dilaton and graviton contributions gives
\begin{equation}
  \mathcal{F}_{12}^{\text{grav} + \Phi} = - \frac{D - 3}{D - 2} \kappa_D^2 m_1
  m_2 - \frac{1}{D - 2} \kappa_D^2 m_1 m_2 = - \kappa_D^2 m_1 m_2 .
\end{equation}

Next, consider the gauge charge. We have $e_{10}^2 = 2 \kappa_{10}^2 /
\alpha'$ for the $D = 10$ Cartan. Thus, $e_D^2 = 2 \kappa_D^2 / \alpha'$ for
the same gauge fields in any $D$, and $O (26 - D)$ rotational invariance fixes
the left-moving gauge kinetic term to be (at fixed $\Lambda$):
\begin{equation}
  \mathcal{L}_L = - \frac{1}{4 e_D^2} \delta_{a b} F^a_{\mu \nu} F^{b \mu \nu} . \qquad \text{Similary,} \qquad
  \mathcal{L}_R = - \frac{1}{4 \tilde{e}_D^2} \delta_{\tilde{a}  \tilde{b}}
  F^{\tilde{a}}_{\mu \nu} F^{\tilde{b} \mu \nu},
\end{equation}
for the right-moving gauge fields, where $\tilde{e}_D^2 = e_D^2$ follows upon
considering the case of vanishing Wilson lines and focusing on the graviphotons
and $B$ photons. Thus,
\begin{equation}
  \mathcal{F}_{12}^{\text{gauge}} = e_D^2  (Q_{1 L} \cdot Q_{2 L} + Q_{1 R}
  \cdot Q_{2 R}) = \frac{2 \kappa_D^2}{\alpha'}  (Q_{1 L} \cdot Q_{2 L} + Q_{1
  R} \cdot Q_{2 R}) .
\end{equation}

Finally, consider the moduli encoded by $\Lambda$. We write
\begin{equation}
  \Lambda = \exp \left[ \left(\begin{array}{cc}
    0 & \lambda\\
    \lambda^{\top} & 0
  \end{array}\right) \right],
\end{equation}
for a small boost away from $\Gamma = \Gamma_0$. Therefore,
\begin{equation}
  Q_L \rightarrow Q_L + \lambda Q_R, \qquad Q_R \rightarrow Q_R +
  \lambda^{\top} Q_L,
\end{equation}
under an infinitesimal boost. Thus,
\begin{equation}
  \frac{\alpha'}{4}  \frac{\partial m^2}{\partial \lambda_{a \tilde{b}}} =
  \frac{\partial}{\partial \lambda_{a \tilde{b}}}  \frac{Q_L^2}{2} =
  \frac{\partial}{\partial \lambda_{a \tilde{b}}}  \frac{Q_R^2}{2} = Q_L^a
  Q_R^{\tilde{b}}\,, \qquad
\text{and so} \qquad
  \frac{\partial m}{\partial \lambda_{a \tilde{b}}} = \frac{Q_L^a
  Q_R^{\tilde{b}}}{\frac{\alpha'}{2} m} .
\end{equation}
Using $O (26 - D) \times O (10 - D)$ rotational invariance, we conclude that
the scalar Lagrangian is
\begin{equation}
  \mathcal{L}= - \frac{1}{2} G_{\lambda \lambda} \delta^{a b}
  \delta^{\tilde{c} \tilde{d}} \nabla \lambda_{a \tilde{b}} \cdot \nabla
  \lambda_{c \tilde{d}},
\end{equation}
for small $\lambda$, therefore the contribution of the moduli $\lambda_{a
\tilde{b}}$ to the mutual force is
\begin{equation}
  \mathcal{F}_{12}^{\lambda} = - G_{\lambda \lambda}^{- 1}  \frac{(Q_{1 L}
  \cdot Q_{2 L})  (Q_{1 R} \cdot Q_{2 R})}{\frac{(\alpha')^2}{4} m_1 m_2} .
\end{equation}
To fix $G_{\lambda \lambda}$, note that turning on a gauge field background
$A^I_m$ corresponds to the boost
\begin{equation}
  \ell_{R m} \rightarrow \ell_{R m} - \sqrt{\frac{\alpha'}{2}} A_m^I \ell_L^I,
  \qquad \ell_L^I \rightarrow \ell_L^I - \sqrt{\frac{\alpha'}{2}} \ell_R^m
  A^I_m,
\end{equation}
where $Q_L = (\ell_{L m}, \ell_L^I)$ and $Q_R = (\ell_{R m})$ for $\ell_{L, R}
= \sqrt{\frac{\alpha'}{2}} k_{L, R}$. We have $G_{A_m^I A_n^J} =
\frac{1}{e_D^2} \delta^{m n} \delta_{I J}$, so $G_{\lambda \lambda} =
\frac{2}{\alpha' e_D^2} = \frac{1}{\kappa_D^2}$.

Putting everything together,
\begin{equation}
  \mathcal{F}_{12} = \frac{2 \kappa_D^2}{\alpha'}  (Q_{1 L} \cdot Q_{2 L} +
  Q_{1 R} \cdot Q_{2 R}) - \kappa_D^2 m_1 m_2 - \kappa_D^2  \frac{(Q_{1 L}
  \cdot Q_{2 L})  (Q_{1 R} \cdot Q_{2 R})}{\frac{(\alpha')^2}{4} m_1 m_2} .
\end{equation}
This miraculously factors:
\begin{equation}
  \mathcal{F}_{12} = - \frac{4 \kappa_D^2}{(\alpha')^2 m_1 m_2}  \left(
  \frac{\alpha'}{2} m_1 m_2 - Q_{1 L} \cdot Q_{2 L} \right) \left(
  \frac{\alpha'}{2} m_1 m_2 - Q_{1 R} \cdot Q_{2 R} \right) .
  \label{eq:forcehet}
\end{equation}
The particles are mutually attractive when both factors in parentheses
are nonvanishing and have the same sign. Taking into account the BPS bound
$m^2 \geq \frac{2}{\alpha'} Q_R^2$, we conclude that the second factor
can never be negative, and mutual repulsion requires either
\begin{equation}
  Q_{1 L} \cdot Q_{2 L} \geq \frac{\alpha'}{2} m_1 m_2,
\end{equation}
or that both particles are BPS, with $Q_{1 R}$ parallel to $Q_{2 R}$ (they are
mutually BPS). In particular, a self-repulsive particle is either BPS
($\tilde{N} = 0$) or has $Q_L^2 \geq \frac{\alpha'}{2} m^2$ ($N=0,1$).

From (\ref{eq:forcehet}) and the BPS bound $m^2 \geq \frac{2}{\alpha'} Q_R^2$, we see also that a particle has non-negative self-force if and only if it satisfies the WGC bound,
\begin{equation}
  \frac{\alpha'}{4} m^2 \leq \frac{1}{2} \max (Q_L^2, Q_R^2) ,
\end{equation}
i.e., it is superextremal. Thus, the RFC bound and the WGC bound are identical.

\bibliographystyle{JHEP}
\bibliography{ref}
\end{document}